\documentclass[a4paper,onecolumn,aps]{revtex4-2}

\newcommand{\eq}[1]{\begin{equation}#1\end{equation}}
\newcommand{\dd}{\mathrm{d}}
\newcommand{\ee}{\mathrm{e}}

\usepackage{amsmath}
\usepackage{amsfonts}
\usepackage{graphics}
\usepackage{graphicx}
\usepackage{psfrag}
\usepackage{color}
\usepackage{colordvi}
\usepackage{bbm}
\usepackage{braket}
\usepackage{physics}

\usepackage{hyperref}

\begin{document}

\title{On the Bisognano-Wichmann entanglement Hamiltonian of nonrelativistic fermions}
\author{Viktor Eisler}

\affiliation{
Institute of Theoretical and Computational Physics, Graz University of Technology,
Petersgasse 16, A-8010 Graz, Austria
}

\begin{abstract}
We study the ground-state entanglement Hamiltonian of free nonrelativistic fermions for semi-infinite domains
in one dimension. This is encoded in the two-point correlations projected onto the subsystem,
an operator that commutes with the linear deformation of the physical Hamiltonian.
The corresponding eigenfunctions are shown to possess the exact same structure both in the continuum as well as on the lattice.
Namely, they are superpositions of the occupied single-particle modes of the total Hamiltonian, weighted by the
inverse of their energy as measured from the Fermi level, and multiplied by an extra phase proportional to the integrated weight.
Using this ansatz, we prove that the Bisognano-Wichmann form of the entanglement Hamiltonian
becomes exact, up to a nonuniversal prefactor that depends on the dispersion for gapped chains.

\end{abstract}

\maketitle

\section{Introduction}

The study of entanglement properties plays a pivotal role in quantum many-body physics \cite{AFOV08,CCD09,ECP10,Laflo16}.
The central object of these investigations is the reduced density matrix (RDM), which encodes all the information
on the entanglement between a subsystem and its remainder. The analogy with the standard setup of statistical mechanics
suggests to write the RDM in an exponential form, and the associated entanglement Hamiltonian (EH) has been the topic of
intensive research \cite{DEFV22}. One of the key questions to address is the characterization of the EH for many-body ground states,
with a particular focus on its locality properties and its relation to the physical Hamiltonian. Beside the pure theoretical interest,
these properties also play a decisive role in novel tomographic protocols developed in quantum simulator experiments
\cite{KBEVZ21,Joshietal23}.

Although extracting the EH for generic many-body systems is an immensely complicated task, one might hope to gain
insight from the study of a related quantum field theory (QFT), that is expected to capture universal features of the model.
In fact, the analog of the EH in algebraic QFT is known as the modular Hamiltonian, and it is associated to an observable
algebra defined on a restricted spacetime region \cite{Haag92,Borchers00}. In particular, considering a wedge region associated
to a semi-infinite subsystem, the seminal result of Bisognano and Wichmann (BW) states, that for a relativistic QFT the modular
Hamiltonian is given by the generator of Lorentz boosts \cite{BW75,BW76}. In one spatial dimension the EH can thus be
constructed as
\eq{
\mathcal{H}_{\textrm{BW}} = \frac{2\pi}{v}\int_{0}^{\infty} x \, T_{00}(x) \, \dd x \, ,
\label{BW}}
where $T_{00}(x)$ is the energy-density component of the stress tensor, and $v$ is the velocity of
excitations, which is required to make the expression dimensionless. In other words, the EH can
be written as a deformation of the physical Hamiltonian by a linear weight function, which could alternatively
be interpreted as a local inverse temperature. While the BW theorem holds for
an arbitrary relativistic QFT, its generalizations to other subsystem geometries require conformal
symmetry \cite{HL82,CHM11,WKZV13,CT16}.

The structure of the EH described by the BW theorem, however, does not only emerge in QFT.
Indeed, the discretized version of \eqref{BW} was found to describe integrable quantum chains in
their gapped phase \cite{PKL99}. This follows from the intimate connection between the RDM and the
corner transfer matrix (CTM) of a corresponding two-dimensional statistical model, which was first introduced
and studied by Baxter \cite{Baxter76,Baxter77,Baxter82}. In fact, it was understood later on, that the generator
of the CTM can be identified as a Lorentz boost operator on the lattice \cite{Tetelman82,Thacker86,IT87,TI88},
which further clarifies the immediate analogy with the BW result.

The CTM method thus yields the EH of integrable chains in a form similar to the original Hamiltonian, but with
couplings that increase linearly from the boundary. In particular, for models that can be mapped into free fermions,
the eigenvalues and eigenvectors of the deformed Hamiltonian were studied in detail \cite{Davies88,TP88,TP89,TP90,ET92}.
For gapped chains one obtains an equidistant spectrum, with a level spacing that goes towards zero at criticality.
In CTM studies, one then usually regularizes the problem by considering a finite chain. However, for such a geometry,
the actual EH is rather given by a sine-deformation of the couplings \cite{EP18}, as predicted by conformal field theory (CFT) \cite{CT16}.
Thus, in order to study the EH at criticality, one needs to work directly in the thermodynamic limit.

The goal of this paper is to show, that the EH for semi-infinite domains can nevertheless be treated on a common
footing for both critical and gapped nonrelativistic free-fermion systems, both in the continuum as well as on the lattice.
We start by considering a one-dimensional Fermi gas in the continuum, and show that the eigenfunctions of the EH are
given by weighted superpositions of the occupied plane wave modes of the physical Hamiltonian. The weight is given by
the inverse of the energy measured from the Fermi surface, and each mode picks up an extra phase proportional to the
integrated weight. The prefactor of this phase is related to the corresponding eigenvalue of the EH, and one recovers an
\emph{exact} BW form \eqref{BW} with the parameter $v$ given by the Fermi velocity. The EH can thus be obtained
directly in the thermodynamic limit, despite the entanglement entropy being ill-defined due to the continuous spectrum. 

In a next step, we extend our discussion to the lattice, considering homogeneous or dimerized hopping chains,
and show that the very same construction holds for the eigenvectors of the EH. While at criticality one obtains a continuous
spectrum, the presence of a gap in the dimerized case induces a quantization as in the CTM studies \cite{PKL99}.
In turn, one finds a BW form with a nonuniversal (mass-dependent) prefactor, which can be related to the properties
of the integrated weight. Finally, we show that the construction can also be applied for a hopping chain with a staggered
chemical potential, yielding an equidistant spectrum which, however, differs from the CTM quantization.

The structure of the paper is as follows. In Section \ref{sec:fg} we consider the EH of the free Fermi gas
on the line, which has a critical Fermi sea ground state. This is followed in Section \ref{sec:hop} by the
study of the analogous lattice problem, the homogeneous hopping chain. The EH for the gapped case
is studied for chains with dimerized hopping and staggered chemical potential in Sections \ref{sec:dim}
and \ref{sec:stag}, respectively. The paper concludes with a discussion in Section \ref{sec:disc},
followed by two appendices presenting some technical details of the calculations.

\section{One-dimensional Fermi gas\label{sec:fg}}

We first consider the free Fermi gas on the infinite line, defined by the single-particle Hamiltonian
\eq{
\hat H = -\frac{1}{2}\frac{\dd^2}{\dd x^2}-\frac{q^2_F}{2}.
\label{HFg}}
The ground state is a Fermi sea, with the plane-wave eigenstates filled up to the 
Fermi momentum $q_F$. The two-point correlations are then given by the sine kernel
\eq{
K(x,x') = \int_{-q_F}^{q_F} \frac{\dd q}{2\pi} \, \ee^{i q (x-x')}=
\frac{\sin q_F(x-x')}{\pi (x-x')},
\label{Ksin}}
which completely characterizes the ground state. We are interested in the entanglement
properties of a semi-infinite bipartition $A=[0,\infty)$, with the reduced density matrix
written in the form
\eq{
\rho_A = \frac{1}{\mathcal{Z}} \ee^{-\hat{\mathcal{H}}},
}
where $\hat{\mathcal{H}}$ is the entanglement Hamiltonian and $\mathcal{Z}$ ensures normalization.
Due to Wick's theorem, the EH is entirely determined by the reduced correlation kernel, which acts as
the integral operator
\eq{
\hat K \psi (x) = \int_{0}^{\infty} \dd x' K(x,x') \psi(x') \, ,
\label{K}}
and is related to the EH as \cite{Peschel03,PE09}
\eq{
\hat K = \frac{1}{\ee^{\hat{\mathcal{H}}}+1}.
\label{KH}}

Hence, in order to find $\hat{\mathcal{H}}$, one needs to solve the eigenvalue problem of the
integral operator $\hat K$. Similarly to the case of a finite interval \cite{SP61}, this task is made easier
by observing that the differential operator 
\eq{
\hat D = -\frac{1}{2}\frac{\dd}{\dd x} x \frac{\dd}{\dd x} - x\frac{q_F^2}{2}
\label{D}}
commutes exactly with the integral operator in \eqref{K}. The calculation of the commutator
$[\hat K, \hat D]=0$ is given in Appendix \ref{app:comm}. Note that the operator \eqref{D} has
precisely the BW form, i.e. it is a simple linear deformation of the physical Hamiltonian in \eqref{HFg}.
In order to solve its eigenvalue problem $\hat D \psi(x)=\lambda \psi(x)$, we first rescale variables as
$y=q_F x$ and $\chi = \lambda/q_F$, such that we arrive at
\eq{
\left(\frac{\dd}{\dd y} y \frac{\dd}{\dd y} + y \right) \psi(y) = -2\chi \, \psi(y).
\label{Dy}}
With a further change of variables $z=-2iy$, we look for a solution of the form $\psi(y) = \ee^{-z/2} \Phi(z)$
such that the differential equation \eqref{Dy} becomes
\eq{
z\frac{\dd^2 \Phi}{\dd z^2} + (1-z)\frac{\dd \Phi}{\dd z} - \left(\frac{1}{2}-i\chi\right) \Phi = 0 \, .
}
This is nothing but the confluent hypergeometric equation with parameters $a=1/2-i\chi$ and $b=1$,
and the solution that is regular at $z=0$ is given by Kummer's function $\Phi(z)=M(a,b,z)$. It can be defined as
a generalized hypergeometric series, however, for our purposes it is more useful to consider
its integral representation \cite{AS64}
\eq{
M(a,b,z) =  \frac{\Gamma(b)}{\Gamma(a)\Gamma(b-a)}
\int_0^1 \dd u \, \ee^{zu} u^{a-1} (1-u)^{b-a-1}.
}
Substituting for the parameters $a,b$ and changing back to the $y$ variable, the solution for
the eigenfunction reads
\eq{
\psi_\chi(y) = \frac{1}{|\Gamma(\frac{1}{2}+i\chi)|^2}
\int_{-1}^{1} \dd p \, \frac{\ee^{iyp} \ee^{i \chi \ln (\frac{1+p}{1-p})}}{\sqrt{1-p^2}},
\label{psiy}}
where we have symmetrized the integral by changing to the variable $p=1-2u$.

Before continuing with our analysis, it is instructive to check that the expression \eqref{psiy}
satisfies the differential equation \eqref{Dy}. Dropping the prefactor and applying the derivatives to the integrand
one finds
\eq{
\int_{-1}^1 \dd p \, [y(1-p^2)+ip] \frac{\ee^{iyp} \ee^{i \chi \ln (\frac{1+p}{1-p})}}{\sqrt{1-p^2}}.
\label{Dpsi}}
One can then integrate by parts in the first term using
\eq{
y \, \ee^{iyp} = -i\frac{\dd}{\dd p}\ee^{iyp}, \qquad
i\frac{\dd}{\dd p}\sqrt{1-p^2} = \frac{-ip}{\sqrt{1-p^2}},
}
such that the second term in the brackets in \eqref{Dpsi} is canceled. The remaining term is
given by the derivative of the logarithmic phase factor, which yields
\eq{
\frac{\dd}{\dd p} \ln (\frac{1+p}{1-p}) = \frac{2}{1-p^2}
\label{dphip}}
such that the $1-p^2$ factor is canceled and one indeed recovers $-2\chi \psi_\chi(y)$.

One can now find a simple interpretation of the result \eqref{psiy}. In fact, the eigenfunction can be thought
of as a wave packet constructed from the momenta within the Fermi sea, with a corresponding amplitude
$1/\sqrt{1-p^2}$. This can be interpreted as a probability inversely proportional to the energy of the given mode,
as measured from the Fermi level. Additionally, each mode acquires an extra phase which, according to \eqref{dphip},
is given by the integral of the weight up to momentum $p$, and has a logarithmic divergence around the Fermi points.
The eigenvalue $\chi$ appears as a factor multiplying this phase, and can assume arbitrary values, i.e. the operator
$\hat D$ has a continuous spectrum. 

\begin{figure}[htb]
\center
\includegraphics[width=0.49\textwidth]{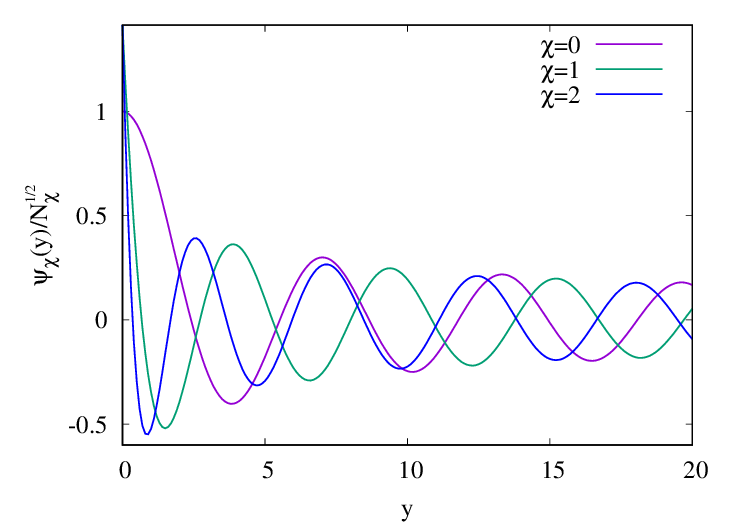}
\includegraphics[width=0.49\textwidth]{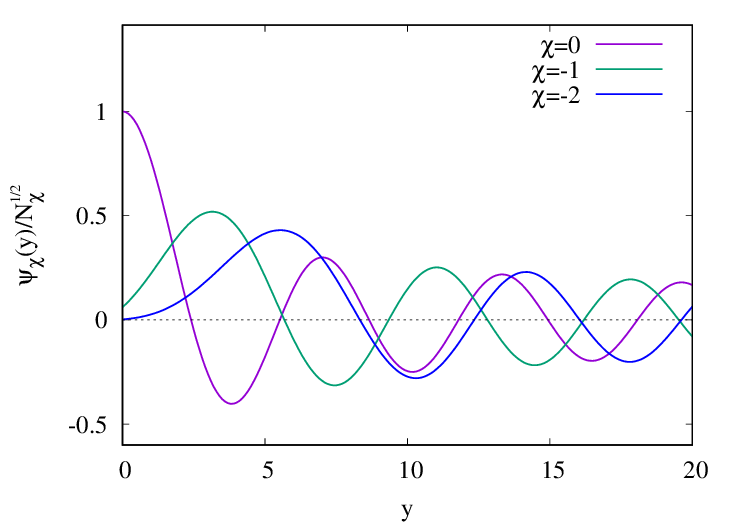}
\caption{Eigenfunctions $\psi_\chi(y)$ of the differential equation \eqref{Dy} for various eigenvalues
$\chi \ge 0$ (left) and $\chi \le 0$ (right). The normalization $N_\chi$ is fixed by \eqref{normpsiy}.}
\label{fig:psiy}
\end{figure}

The eigenfunctions are shown in Fig.~\ref{fig:psiy} for various eigenvalues $\chi$. Note that, due to the continuous spectrum,
the $\psi_\chi(y)$ are not normalizable, and the factor in front of the integral in \eqref{psiy} sets $\psi_\chi(0)=1$.
However, in Fig.~\ref{fig:psiy} we adopt a different normalization, that comes from the requirement of orthonormality and
will be derived later. One observes that, for $\chi>0$, the oscillations increase and $\psi_\chi(y)$ becomes more and more
peaked around $y=0$, whereas for $\chi<0$ the eigenfunction is pushed away from the boundary.

To find the asymptotics for $y \gg 1$, it is useful to first introduce new variables $p=\tanh(z)$, in order to remove the
divergent phase factor in \eqref{psiy}. One then has
\eq{
\psi_\chi(y) = \frac{\cosh(\pi \chi)}{\pi}
\int_{-\infty}^{\infty} \dd z \, \frac{\ee^{iy \tanh(z)} \ee^{i 2\chi z}}{\cosh(z)},
\label{psiz}}
where we used the properties of the $\Gamma$ function for imaginary arguments. One can now apply a stationary phase
approximation, which yields the condition
\eq{
\frac{y}{\cosh^2{z_0}} + 2\chi =0 \, , \qquad
\tanh(z_0) = \sqrt{1-\frac{2|\chi|}{y}}.
}
Note that the stationary points $ \pm z_0$ exist only for $\chi < 0$, and one finds, up to normalization, the approximation
\eq{
\psi_\chi(y) \propto
\sqrt{\frac{\pi}{y\sqrt{1-\frac{2|\chi|}{y}}}}
\exp\left[iy \sqrt{1-\frac{2|\chi|}{y}}+i2\chi \, \mathrm{atanh}\sqrt{1-\frac{2|\chi|}{y}}-i\pi/4\right]
+ \mathrm{c.c.}
}
In particular, for $y \gg 2|\chi|$ the second phase factor can be expanded and one finds $\chi \ln(2y/|\chi|)$, such that
the frequency of the oscillations increases logarithmically.

The commutation property $[\hat K, \hat D]=0$ ensures, that $\psi_\chi(y)$ are also the eigenfunctions of the sine kernel
on the half-line, it thus remains to evaluate the corresponding eigenvalue. For this purpose, we apply the substitution
used in \eqref{psiz} also in the sine kernel to find
\eq{
K(y,y')=
\int_{-\infty}^{\infty} \frac{\dd z}{2\pi} \, \frac{\ee^{i (y-y') \tanh(z)} }{\cosh^2(z)} \, .
}
Applying the integral operator to $\psi_\chi(y')$, the $y'$ integral has to carried out on the half line,
one thus simply obtains the Fourier transform of the step function $\Theta(y')$, which is given by
\eq{
\int_{-\infty}^{\infty } \dd y' \, \Theta(y') \, \ee^{iQy'} = \lim_{\epsilon\to 0^+} \frac{i}{Q+i\epsilon}.
}
Ignoring the normalization factor, the function $\hat K \psi_\chi(y)$ thus has the integral representation
\eq{
\int_{-\infty}^{\infty}  \dd z \, \frac{\ee^{i y \tanh(z)} }{\cosh^2(z)}
\lim_{\epsilon\to 0^+} \oint_{\gamma} \frac{\dd z'}{2\pi} \,
\frac{i}{\tanh(z')-\tanh(z)+i\epsilon} \frac{\ee^{i 2\chi z'}}{\cosh(z')},
\label{contint}}
where we have extended the $z'$ integration to a contour $\gamma$ on the complex plane.
For $\chi > 0$, the contour is chosen as an infinitely large semi-circle on the upper half-plane,
whereas for $\chi<0$ the contour must be closed on the lower half-plane. With this choice,
the contribution of the integral on the arc vanishes, and one only needs to evaluate the
residues at the poles which lie at
\eq{
z' = z + i n \pi - \frac{i \epsilon}{1-\tanh^2(z)},
\label{poles}}
where $n$ is an integer.

Let us first consider $\chi>0$, such that the poles that lie within the contour
correspond to $n=1,2,\dots$, and their residue is given by $\cosh^2(z)$. At the same time, the factor
in the denominator becomes $\cosh(z')=(-1)^n \cosh(z)$, such that we have for the contour integral
\eq{
-\lim_{\epsilon\to 0} \oint_{\Gamma} \frac{\dd z'}{2\pi i} \,
\frac{1}{\tanh(z')-\tanh(z)+i\epsilon} \frac{\ee^{i 2\chi z'}}{\cosh(z')}=
-\ee^{i 2\chi z} \cosh(z)\sum_{n=1}^{\infty} (-1)^n \ee^{-n 2\pi\chi} \, .
\label{contint2}
}
Plugging the result back into \eqref{contint}, it is easy to see that the $z$ dependent terms reproduce
the eigenfunction $\psi_\chi(y)$, while the corresponding eigenvalue is given by the sum. Using
the formula for the geometric series, one arrives at
\eq{
\hat K \psi(y) = \frac{1}{\ee^{2\pi\chi}+1} \psi(y) \, .
\label{Kpsi}}
The case $\chi<0$ is very similar, but now the poles on the lower half plane with $n=0,-1, \dots$ contribute
in the sum in \eqref{contint2}, while the sign factor is absorbed by changing the direction of the contour integration,
such that the sum delivers the very same result \eqref{Kpsi}.
One thus finds, that the eigenvalue of the sine kernel is simply given by the Fermi function with an argument $2\pi \chi$.
Comparing with \eqref{KH} and restoring the length scales $y=x q_F$ and $\chi = \lambda/q_F$, one finds
\eq{
\mathcal{\hat H} = \frac{2\pi}{q_F}\hat D,
}
and thus the BW theorem \eqref{BW} holds exactly. In fact, despite the infrared divergence of the entanglement
entropy due to the continuum spectrum, the EH remains perfectly well defined.

To conclude this section, we check the orthonormality of the eigenfunctions. To this end we need to evaluate
\eq{
\int_0^{\infty} \dd y \, \psi^*_\chi(y) \psi_{\chi'}(y)=
\frac{\cosh^2(\pi \chi)}{\pi^2}
\int_{-\infty}^{\infty} \dd z \, \frac{2\pi}{\ee^{2\pi\chi}+1} \ee^{i 2(\chi'-\chi) z},
}
where we have carried out the same contour integration as before. In turn, one arrives at
\eq{
\int_0^{\infty} \dd y \, \psi^*_\chi(y) \psi_{\chi'}(y)= N_\chi \delta(\chi-\chi'), \qquad
N_\chi = \frac{1 + \ee^{-2\pi \chi}}{2} \, ,
\label{normpsiy}}
and to ensure the correct delta function normalization between the eigenfunctions, $\psi_\chi(y)$ must be
multiplied by a factor $N_\chi^{-1/2}$, which was adopted in Fig. \ref{fig:psiy}.

\section{Homogeneous hopping chain\label{sec:hop}}

We now move on to consider free fermions on an infinite chain, given by a hopping model of the form
\eq{
\hat H = -\frac{1}{2}\sum_{n=-\infty}^{\infty} (c^\dag_n c_{n+1} + c^\dag_{n+1} c_{n})
+ \cos q_F \sum_{n=-\infty}^{\infty} c^\dag_n c_n \, .
}
The chemical potential $\mu=-\cos q_F$ is chosen such that the Fermi momentum is given by $q_F$.
The reduced correlation matrix $C_{m,n}=\langle c_m^\dag c_n \rangle$ with $m,n \ge 1$ is given by the
discrete version of the sine kernel
\eq{
C_{m,n} = \frac{\sin q_F(m-n)}{\pi (m-n)},
\label{C}}
and encodes all the information on entanglement. In particular, the EH is given by the relation \cite{Peschel03,PE09}
\eq{
\mathcal{H} = \sum_{m,n \ge 1} H_{m,n}c_m^\dag c_n \, , \qquad
C = \frac{1}{\ee^{H}+1}.
\label{CH}}

We thus have to treat the eigenvalue problem of the discrete sine kernel \eqref{C} on the half-infinite chain.
Similarly to the continuum case, this can be simplified by finding a commuting operator with a much simpler
structure.  Indeed, analogously to the case of an interval \cite{Slepian78,Peschel04},
one can show (see Appendix \ref{app:comm}) that the tridiagonal matrix
\eq{
T_{m,n} =  t_m\, \delta_{m+1,n} + t_{m-1} \, \delta_{m-1,n} + d_m \delta_{m,n} \,
\label{T}}
with entries defined by 
\eq{
t_m = -\frac{1}{2}m, \qquad
d_m = \cos q_F (m-1/2),
\label{Thom}}
commutes exactly with the correlation matrix, $[C,T]=0$. Just as in the continuum case,
the commuting operator has precisely the BW form, with the linear deformation being slightly shifted
for the on-site $d_m$ and hopping terms $t_m$, respectively.

In analogy with the continuum case, we try the following ansatz for the eigenvector
\eq{
\psi_\lambda(m) =
\int_{-q_F}^{q_F} \frac{\dd q}{2\pi} \,
\frac{\ee^{iq(m-1/2)} \ee^{i \lambda \varphi_q }}{\sqrt{\cos q-\cos q_F}},
\label{psim}}
where the denominator contains again the square root of the energies measured
from the Fermi level. The variable $m-1/2$ in the first phase factor is suggested by the reflection
symmetry $m \to 1-m$ of the problem, which should translate to the eigenvalue $-\lambda$,
whereas the phase $\varphi_q$ is yet to be determined. Multiplying with the tridiagonal matrix,
the integrand of the vector $-2T \psi_\lambda$ reads
\eq{
[(2m-1) (\cos q-\cos q_F) + i \sin q]
\frac{\ee^{iq(m-1/2)} \ee^{i \lambda \varphi_q}}{\sqrt{\cos q-\cos q_F}}.
\label{Tpsi}}
One can then integrate by parts in the first term using
\eq{
(2m-1) \ee^{iq(m-1/2)} = -2i\frac{\dd}{\dd q} \ee^{iq(m-1/2)},\qquad
2i\frac{\dd}{\dd q} \sqrt{\cos q-\cos q_F} = 
\frac{-i\sin q}{\sqrt{\cos q-\cos q_F}},
}
such that the derivative of the square root cancels the second term in \eqref{Tpsi},
while the boundary term vanishes automatically. One has thus the requirement
\eq{
\frac{\dd \varphi_q}{\dd q_{\phantom{q}}} = \frac{1}{\cos q - \cos q_F},
\label{dphiq}}
which yields $-2T \psi_\lambda = -2\lambda\psi_\lambda$. Hence, in complete analogy to the
continuum case \eqref{dphip}, the extra phase is given by the integral of the inverse energy difference,
which gives
\eq{
\varphi_q = \frac{1}{\sin q_F}
\ln (\frac{\tan \frac{q_F}{2} + \tan \frac{q}{2}}{\tan \frac{q_F}{2}-\tan \frac{q}{2}}).
}
Note that the lower limit of the integral was set to $q=0$, which ensures that the phase is odd, $\varphi_{-q}=\varphi_q$,
and thus the eigenvector in \eqref{psim} is real.

The next step is to substitute variables $\tanh(z)  = \tan(q/2)/\tan(q_F/2)$, to cure the divergence of the phase factor.
Indeed, this gives $\varphi_q = 2z/\sin q_F$, and the inverse transformation and its Jacobian read
\eq{
q(z)= 2 \atan[\tan(q_F/2)\tanh (z)],\qquad
\frac{\dd q}{\dd z} =\frac{2 \sin q_F}{\cos q_F + \cosh (2z)},
\label{qz}}
while the energy difference can be expressed as
\eq{
\cos q - \cos q_F = 
\frac{2\sin^2(q_F/2)}{\cosh^2(z)[1+\tan^2(q_F/2)\tanh^2(z)]}=
\frac{\sin^2 q_F}{\cos q_F + \cosh(2z)}.
\label{cosqdiff}}
Putting everything together, the eigenvector can be written as
\eq{
\psi_\lambda(m) =
\int_{-\infty}^{\infty} \frac{\dd z}{\pi} \,
\frac{\ee^{iq(z)(m-1/2)} \ee^{i 2\lambda z/\sin q_F}}{\sqrt{\cos q_F + \cosh (2z)}}.
}

With the integral representation at hand, one could proceed to evaluate the corresponding
eigenvalue of $C$. We first rewrite its matrix elements using the transformation \eqref{qz} as
\eq{
C_{m,n} = \int_{-\infty}^{\infty} \frac{\dd z}{2\pi} \, \frac{\dd q}{\dd z} \, \ee^{i q(z) (m-n)}.
}
To evaluate the matrix multiplication $C \psi_\lambda$, we use the identity
\eq{
\sum_{n=1}^{\infty}\ee^{i Q (n-1/2)} =
\lim_{\epsilon\to 0^+} \frac{i}{2 \sin (\frac{Q}{2}+i\epsilon)},
}
which leads us to the contour integral
\eq{
\int_{-\infty}^{\infty} \frac{\dd z}{2\pi} \, \frac{\dd q}{\dd z} \, \ee^{i q(z)(m-1/2)}
\lim_{\epsilon\to 0^+}\oint_{\gamma} \frac{\dd z'}{2\pi} \frac{\dd q}{\dd z'}
\frac{i}{2 \sin (\frac{q(z')-q(z)}{2}+i\epsilon)}
\frac{\ee^{i 2\lambda z'/\sin q_F}}{\sqrt{\cos q(z')-\cos q_F}}.
}
The poles of the integrand are the same \eqref{poles} as in the previous section, 
and for $\lambda>0$ ($\lambda<0$) we can close the contour on the upper (lower) half-plane.
The residues are simply given by $(\frac{\dd q}{\dd z})^{-1}$ which cancels with the Jacobian. Due to the
presence of the square root in the denominator, some care is needed when extending
the weight factor to the complex plane, in order to avoid branch cuts. Indeed, here one
should use the first expression in \eqref{cosqdiff}, which implies
\eq{
\sqrt{\cos q(z')-\cos q_F} = (-1)^n \sqrt{\cos q(z)-\cos q_F} \, .
}
The structure is thus exactly the same as in \eqref{contint}, and yields immediately the relations.
\eq{
C \psi_\lambda = \frac{1}{\ee^{\frac{2\pi \lambda}{\sin q_F}}+1} \psi_\lambda \, ,
\qquad
H = \frac{2\pi}{\sin q_F} T \, .
\label{Cpsi}}
In other words, the BW form \eqref{BW} is exact even on the lattice and at arbitrary fillings, setting $v=\sin q_F$.

\begin{figure}[htb]
\center
\includegraphics[width=0.49\textwidth]{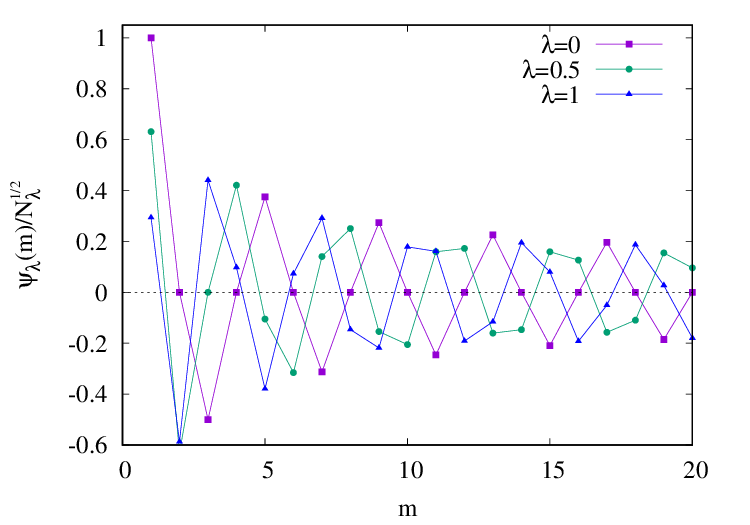}
\includegraphics[width=0.49\textwidth]{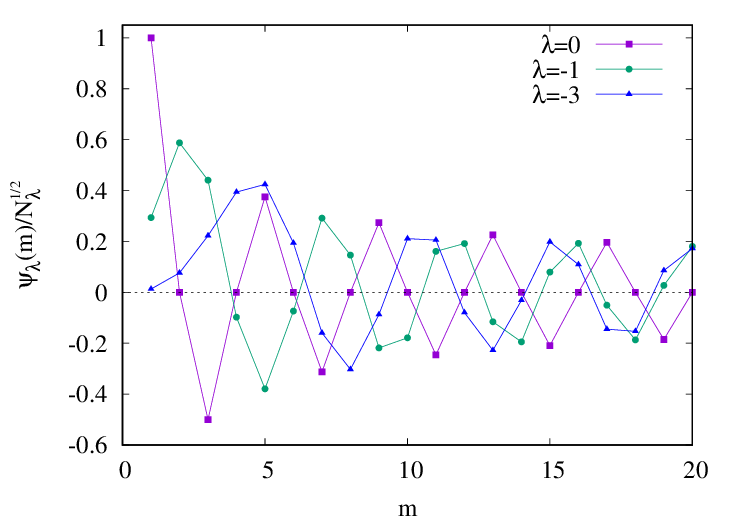}
\caption{Eigenvectors $\psi_\lambda(m)$ of the tridiagonal matrix \eqref{T} for various eigenvalues
$\lambda \ge 0$ (left) and $\lambda \le 0$ (right). The normalization $N_\lambda$ is fixed by \eqref{normpsim}.}
\label{fig:psim}
\end{figure}

The normalization of the eigenvectors can be obtained analogously to the previous section
\eq{
\sum_{m=1}^{\infty} \psi^*_\lambda(m) \psi_{\lambda'}(m)= N_\lambda \delta(\lambda-\lambda'), \qquad
N_\lambda = \frac{1}{\ee^{\frac{2\pi \lambda}{\sin q_F}}+1} \, .
\label{normpsim}}
Note that the difference with respect to \eqref{normpsim} comes from the prefactor in \eqref{psiy},
which is missing from the ansatz \eqref{psim}. The properly normalized eigenvectors are shown
in Fig.~\ref{fig:psim}, and follow the same pattern as observed in Fig.~\ref{fig:psiy} for the continuous case.
Namely, for $\lambda>0$ the eigenvectors are attracted, while for $\lambda<0$ they are repelled from
the boundary. Note that for larger $\lambda>0$ the structure becomes quickly rather irregular,
which is due to the discrete sampling from an increasingly oscillatory function.

Finally, we point out a remarkable connection of our ansatz \eqref{psim} to orthogonal polynomials.
Indeed, it was already observed in earlier studies of corner transfer matrix spectra \cite{TP88,TP90},
that the eigenvectors of $T$ are given by the Meixner-Pollaczek polynomials.
It is rooted in the fact, that $\psi_\lambda(m)$ satisfy a three-term recurrence relation.
In particular, the Meixner-Pollaczek polynomials are defined by the recurrence relation \cite{Koekoek10}
\eq{
(n+1)P^{(\alpha)}_{n+1}(x;\phi) + (n+2\alpha-1)P^{(\alpha)}_{n-1}(x;\phi) =
2(x \sin \phi + (n+\alpha) \cos \phi) P^{(\alpha)}_{n}(x;\phi) \, .
}
It is easy to see that setting $\alpha=1/2$, $\phi=q_F$ and $x=-\lambda/\sin q_F$, one obtains
precisely the eigenvalue equation of the tridiagonal matrix with elements \eqref{Thom} under
the identification $P^{(1/2)}_{n}(x;q_F) \sim \psi_\lambda(n+1)$ for $n=0,1,\dots$.
In fact, it turns out that there is a proportionality factor between them, which has to be fixed by
comparing the orthogonality relations satisfied by the Meixner-Pollaczek
polynomials, which is carried out in Appendix \ref{app:MP}. 

\section{Dimerized chain\label{sec:dim}}

The dimerized (or SSH) chain is described by a hopping model with alternating amplitudes
\eq{
\hat H = - \sum_{n} \left( \frac{1+\delta}{2} \, c_{2n-1}^\dag c_{2n} + 
\frac{1-\delta}{2} \, c_{2n}^\dag c_{2n+1} + \mathrm{h.c.} \right)
\label{h}}
Since the chain has only two-site shift invariance, a Fourier transformation in terms of the
sublattice momentum must be followed by a rotation within the cell degrees of freedom
in order to diagonalize the model. This is most easily expressed in terms of the halved
sublattice momentum, which varies within the reduced Brillouin zone $q\in [-\pi/2,\pi/2]$.
In particular, we introduce new fermion operators $\alpha_q$ and $\beta_q$ via the
transformation
\eq{
c_{2n-1} =\frac{1}{\sqrt{2N}} \sum_q \ee^{iq(2n-1)}\ee^{-i\theta_q/2}(\alpha_q + \beta_q) \, ,
\qquad
c_{2n} =\frac{1}{\sqrt{2N}} \sum_q \ee^{iq 2n}\ee^{i\theta_q/2}(\alpha_q - \beta_q) \, .
\label{aqbq}}
It is then easy to see, that the phase factor must be chosen as
\eq{
\ee^{i\theta_q} = \frac{\cos q - i\delta \sin q}{\omega_q}, \qquad
\omega_q = \sqrt{\cos^2 q+\delta^2 \sin^2 q} \, ,
}
in order to bring the Hamiltonian into the diagonal  from
\eq{
\hat H = -\sum_q \omega_q (\alpha_q^\dag \alpha_q - \beta_q^\dag \beta_q) \, .
\label{Hdimdiag}}
Note that, for simplicity, we have formulated the transformations above on a periodic chain of
$2N$ sites. However, we are interested in the limit $N\to\infty$ where the momentum variable $q$
becomes continuous. 

The Hamiltonian \eqref{Hdimdiag} is characterized by a band structure with a gap of size $2|\delta|$,
and the ground state is given by filling only the lower band, i.e. $\langle \alpha_q^\dag \alpha_q \rangle=1$
and $\langle \beta_q^\dag \beta_q \rangle=0$. The correlation matrix still has a checkerboard structure,
with the nonvanishing entries given by
\eq{
C_{2m-1,2n}=\int_{-\pi/2}^{\pi/2} \frac{\dd q}{2\pi} \, \ee^{i q r} \ee^{i\theta_q}, \qquad
C_{2m,2n+1}=\int_{-\pi/2}^{\pi/2} \frac{\dd q}{2\pi} \, \ee^{i q r} \ee^{-i\theta_q},
}
where $r=2n-2m+1$, and $C_{2m-1,2n-1}=C_{2m,2n}=\delta_{m,n}/2$. It is a simple exercise to show
(see Appendix \ref{app:comm}), that one can again define a commuting tridiagonal matrix with the hopping
amplitudes given by
\eq{
t_{2m-1} = -\frac{1+\delta}{2}(2m-1), \qquad
t_{2m} = -\frac{1-\delta}{2}2m, \qquad
\label{tmdim}}
while the diagonal terms are zero, $d_m = 0$.

Inspired by the results of the previous sections and the structure of the transformation \eqref{aqbq},
we propose the ansatz
\eq{
\psi_\lambda(m) =
\int_{-\pi/2}^{\pi/2} \frac{\dd q}{2\pi} \,
\frac{\ee^{iq(m-1/2)} \ee^{i \lambda\varphi_q}\ee^{i(-1)^m\theta_q/2}}{\sqrt{\omega_q}}=
\int_{-\pi/2}^{\pi/2} \frac{\dd q}{2\pi} \,
\frac{\ee^{iq(m-1/2)} \ee^{i \lambda \varphi_q}}{\sqrt{\cos q + (-1)^m i \delta \sin q}},
\label{psimdim}}
where we incorporated the additional phase factors into the wavefunction. Multiplying by $-2T$
using \eqref{tmdim} and assuming $m$ odd, one obtains for the integrand
\eq{
\left[(2m-1)(\cos q + i \delta \sin q) + (i \sin q + \delta\cos q)\right]
\frac{\ee^{iq(m-1/2)} \ee^{i \lambda \varphi_q}}{\sqrt{\cos q + i \delta \sin q}}.
\label{Tpsidim}}
Rewriting the factor $2m-1$ as a $q$-derivative and integrating by parts, one has
for the derivative of the square root
\eq{
2i\frac{\dd}{\dd q} \left[\sqrt{\cos q + i \delta \sin q}\right]=
\frac{-i\sin q - \delta \cos q}{\sqrt{\cos q + i \delta \sin q}},
}
which again exactly cancels the second term in \eqref{Tpsidim}. To obtain the eigenvalue $-2\lambda$,
the extra phase has to satisfy
\eq{
\frac{\dd \varphi_q}{\dd q_{\phantom{q}}}=
\frac{1}{\sqrt{\cos^2 q + \delta^2 \sin^2 q}},
}
which can be integrated as
\eq{
\varphi_q = F(q,\delta'),
\qquad
\delta' = \sqrt{1-\delta^2},
}
in terms of the incomplete elliptic integral of the first kind, and we introduced
the complementary modulus $\delta'$.

In sharp contrast to the critical case, however, the boundary contribution to the partial integral
does not vanish automatically, due to the presence of the gap. It is easy to see that the vanishing
of this term requires
\eq{
\ee^{i\frac{\pi}{2}(m-1/2)} \sqrt{i \delta } \ee^{i \lambda\varphi_{\pi/2}}=
\ee^{-i\frac{\pi}{2}(m-1/2)} \sqrt{-i \delta} \ee^{-i \lambda\varphi_{\pi/2}},
\label{boundary}}
and setting $m=2\ell+1$ this further implies
\eq{
\lambda_\ell =\frac{\pi}{2K(\delta')}
\begin{cases}
2\ell + 1 & \textrm{if $\delta>0$} \\ 
2\ell & \textrm{if $\delta<0$} \\ 
\end{cases}, \qquad
\ell = 0, \pm1, \pm 2,\dots
\label{lamdim}}
where $K(\delta')$ denotes the complete elliptic integral.
Hence, it is precisely this extra requirement that imposes the quantization of the eigenvalues in the gapped case.
The levels are equidistant and their spacing decreases approaching the critical point $\delta \to 0$, where $K(\delta')\to\infty$,
signaling the transition to a continuum spectrum. Note also that the spectrum is particle-hole symmetric, and one
can easily verify that the eigenvectors satisfy the property $\psi_{-\lambda}(m) = (-1)^m \psi_\lambda(m)$.
Furthermore, one can check that the derivation for $m$ even yields the very same results.

\begin{figure}[htb]
\center
\includegraphics[width=0.35\textwidth]{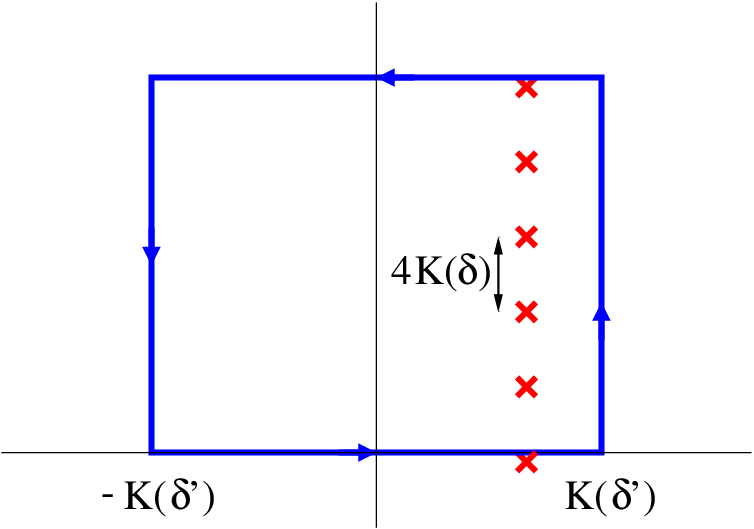}
\caption{Rectangular integration contour (blue) for the eigenvector \eqref{cintdim}, with the upper side taken to infinity.
The red crosses indicate the poles $u' = u + in \, 4K(\delta)-i\epsilon$ that appear when multiplying with the matrix $C$.}
\label{fig:contour}
\end{figure}

We now move to the calculation of the corresponding eigenvalues of $C$. As in the previous cases, we
introduce the phase $u=\varphi_q$ as a new variable, which allows us to extend the integral to the complex
plane with vanishing contributions at infinity. The inverse transformation and its Jacobian read
\eq{
q(u) = \mathrm{am}(u,\delta'), \qquad
\frac{\dd q}{\dd u} = \mathrm{dn}(u,\delta'),
}
where $\mathrm{am}$ and $\mathrm{dn}$ denote the Jacobi and delta amplitude, respectively,
and one obtains the contour integral
\eq{
\psi_\lambda(m) =
\oint_{\gamma} \frac{\dd u}{2\pi} \, \mathrm{dn}(u,\delta')
\frac{\ee^{i \, \mathrm{am}(u,\delta')(m-1/2)} \ee^{i \lambda u}}{\sqrt{\mathrm{cn}(u,\delta') + i(-1)^m \delta \, \mathrm{sn}(u,\delta')}},
\label{cintdim}}
where $\mathrm{sn}$ and $\mathrm{cn}$ are the elliptic sine and cosine, respectively. The contour
$\gamma$ has a rectangular shape as depicted in Fig. \ref{fig:contour} for $\lambda>0$. Indeed, the main
difference w.r.t. the critical case is that the domain on the real axis is a finite interval $[-K(\delta'),K(\delta')]$.
At its endpoints the integrand has the same value due to the quantization condition \eqref{boundary}.
One can show that this remains true along the vertical lines in Fig. \ref{fig:contour}, such that their
contributions cancel, and one can close the contour at infinity. For $\lambda<0$, the contour must be drawn
on the lower half plane.

From this point, the calculation is completely analogous to the critical case. The poles will be given by
the condition $\mathrm{am}(u',\delta')=\mathrm{am}(u,\delta')$ which, using the periodicity properties
of the Jacobi amplitude along the imaginary direction, yields $u' = u + in \, 4K(\delta)$. The square root
in the denominator of \eqref{cintdim} yields again a factor $(-1)^n$, and the infinite sum over the poles
reproduces the Fermi function. One thus obtains for the EH and its spectrum
\eq{
H = 4K(\delta) T, \qquad
\varepsilon_\ell =2\pi \frac{K(\delta)}{K(\delta')}
\begin{cases}
2\ell + 1 & \textrm{if $\delta>0$} \\ 
2\ell & \textrm{if $\delta<0$} \\ 
\end{cases}.
}
One should note that the result seemingly differs from the one obtained via the duality
with two interlaced transverse Ising chains \cite{EDGTP20}, which gives $\pi K(k')/K(k)$
for the halved spacing of the spectrum with $k=\frac{1-|\delta|}{1+|\delta|}$. However, one can show
using elliptic integral identities, that the two results are actually identical.

\begin{figure}[htb]
\center
\includegraphics[width=0.49\textwidth]{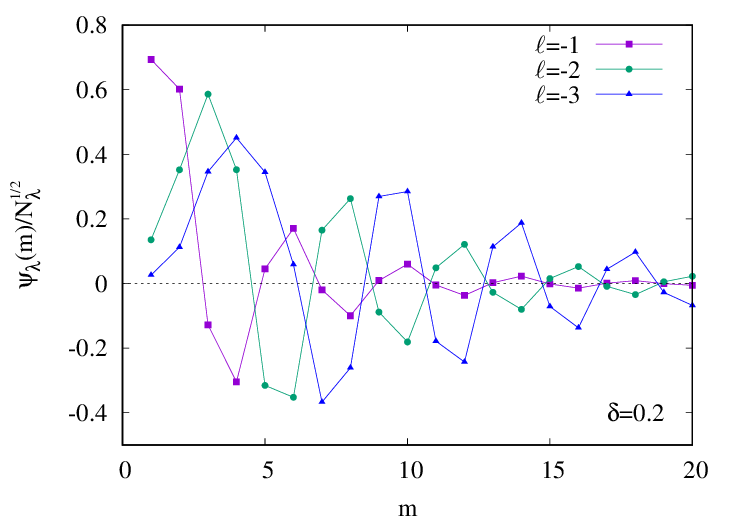}
\includegraphics[width=0.49\textwidth]{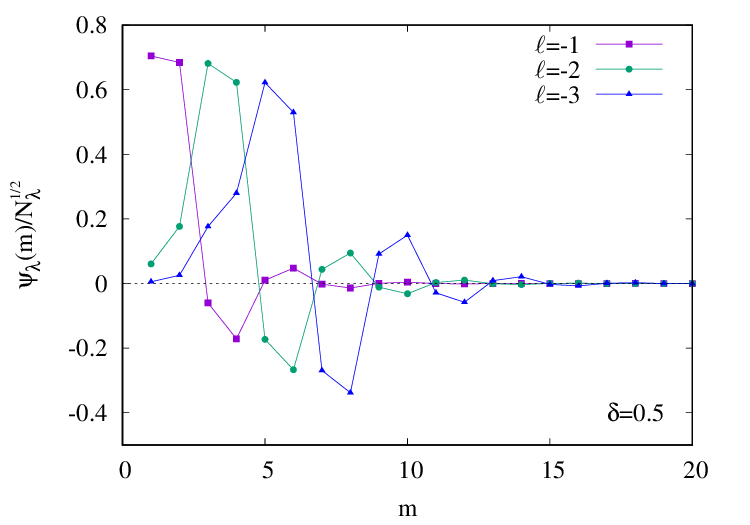}
\caption{Eigenvectors $\psi_\lambda(m)$ of the tridiagonal matrix of the dimerized chain with elements \eqref{tmdim},
for various eigenvalues \eqref{lamdim} and two different dimerizations $\delta=0.2$ (left) and $\delta=0.5$ (right).
The normalization $N_\lambda$ is fixed by \eqref{normpsimdim}.}
\label{fig:psimdim}
\end{figure}

Finally, the normalization of the eigenvectors can also be obtained using the contour
integral representation as
\eq{
\sum_{m=1}^{\infty} \psi^*_\lambda(m) \psi_{\lambda'}(m)= N_\lambda \delta_{\lambda,\lambda'}, \qquad
N_\lambda = \frac{1}{\ee^{4K(\delta) \lambda}+1} \frac{K(\delta')}{\pi}\, .
\label{normpsimdim}}
The factor multiplying the Fermi function is simply the size of the integration domain
on the real axis divided by $2\pi$, and diverges as $\delta \to 0$. The eigenvectors for
two different dimerizations are shown in Fig.~\ref{fig:psimdim}. One can clearly see that
their amplitude decays much faster as in the critical case, especially for larger values of the gap.
Note that we have shown only eigenvalues with $\lambda<0$, as those with $-\lambda$
are simply multiplied by a factor $(-1)^m$.

\section{Hopping chain with staggered potential\label{sec:stag}}

As a final example, let us consider the hopping chain in a staggered chemical potential
\eq{
\hat H = -\frac{1}{2}\sum_{n} (c^\dag_n c_{n+1} + c^\dag_{n+1} c_{n})
+ \mu \sum_{n} (-1)^n c^\dag_n c_n \, .
\label{Hstag}}
The unit cell contains again two sites and one can proceed as in the previous section.
Using the reduced sublattice momentum $q$ and introducing new operators via
\eq{
\begin{split}
&c_{2m-1} =\frac{1}{\sqrt{2N}} \sum_q \ee^{iq(2m-1)}
\Big(\sqrt{1+\frac{\mu}{\Omega_q}}\alpha_q + 
\sqrt{1-\frac{\mu}{\Omega_q}} \ee^{iq} \beta_q \Big) , \\
&c_{2m} =\frac{1}{\sqrt{2N}} \sum_q \ee^{iq 2m}
\Big(\sqrt{1-\frac{\mu}{\Omega_q}}\alpha_q - 
\sqrt{1+\frac{\mu}{\Omega_q}} \ee^{iq} \beta_q \Big) ,
\end{split}
\label{aqbqstag}}
one arrives at the diagonal form of the Hamiltonian and corresponding dispersion relation
\eq{
H = -\sum_q \Omega_q (\alpha_q^\dag \alpha_q - \beta_q^\dag \beta_q) \, ,
\qquad
\Omega_q = \sqrt{\cos^2 q+\mu^2} \, .
\label{dispstag}}
The ground state is again given by the filled band of $\alpha_q$ fermions, and in appendix
\ref{app:comm} we show that the corresponding correlation matrix commutes with the
tridiagonal matrix defined by
\eq{
t_m = -\frac{1}{2}m, \qquad d_m = (-1)^{m}\mu (m-1/2) \, .
}

Comparing with \eqref{aqbq}, we see that the transformation \eqref{aqbqstag} now
includes, instead of a phase, a different weight factor for even and odd sites.
Incorporating this extra weight, we shall use the ansatz for the eigenvectors
\eq{
\psi_\lambda(m) = \int_{-\pi/2}^{\pi/2} \frac{\dd q}{2\pi} \,
\ee^{iq(m-1/2)} \ee^{i \lambda\varphi_q}\frac{\sqrt{\Omega_q-(-1)^m\mu}}{\Omega_q}.
\label{psimstag}}
Multiplying with $-2T$ and using the expression of the dispersion \eqref{dispstag}, one obtains the integrand
\eq{
\left[(2m-1)\sqrt{\Omega_q-(-1)^m\mu}
+i \sin q
\frac{\sqrt{\Omega_q+(-1)^m\mu}}{\Omega_q} \right]\ee^{iq(m-1/2)}\ee^{i \lambda\varphi_q}.
}
Integrating by parts in the first term, one finds again a cancellation with the second term, and 
the phase must satisfy $\frac{\dd \varphi_q}{\dd q_{\phantom{q}}} = \Omega^{-1}_q$,
which can be integrated as
\eq{
\varphi_q =  \kappa \, F(q,\kappa), \qquad
\kappa = \frac{1}{\sqrt{1+\mu^2}}.
}
Furthermore, the boundary contribution can be assessed by noting that $\Omega_{\pm\pi/2}=|\mu|$,
and thus for $\mu>0$ ($\mu < 0$) one needs to consider only the case $m=2\ell+1$ ($m=2\ell$).
The requirement is that the phase factor $\ee^{i\frac{\pi}{2}(m-1/2)}\ee^{i \lambda\varphi_{\pi/2}}$ be real,
which yields
\eq{
\lambda_\ell = \frac{\pi}{2\kappa \, K(\kappa)}
\begin{cases}
2\ell-\frac{1}{2} & \textrm{if $\mu>0$}\\
2\ell+\frac{1}{2} & \textrm{if $\mu<0$}
\end{cases}, \qquad
\ell = 0, \pm1, \pm 2,\dots
\label{lamstag}}
Note that the difference in the quantization w.r.t. \eqref{lamdim} is due to the missing
phase factor in the transformation \eqref{aqbqstag}. In particular, \eqref{lamstag} breaks
the particle-hole symmetry of the spectrum, which transforms as $\lambda_\ell \to -\lambda_{-\ell}$
under $\mu \to -\mu$.

\begin{figure}[htb]
\center
\includegraphics[width=0.55\textwidth]{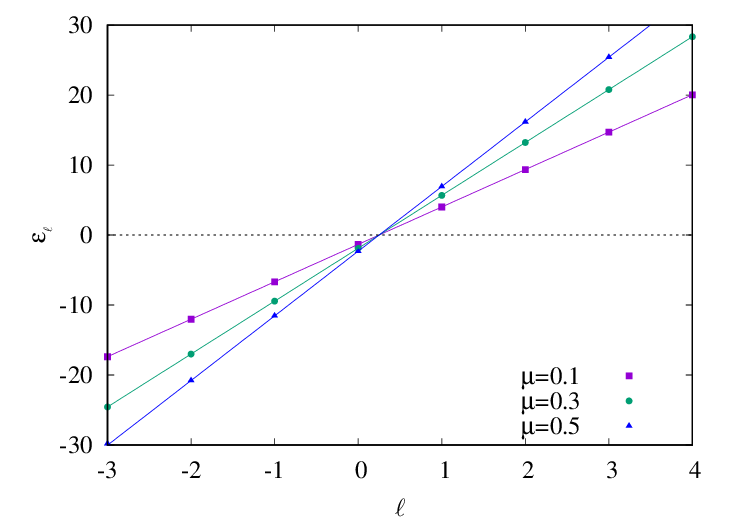}
\caption{Single-particle spectra (symbols) of the EH of a hopping chain with a staggered field,
for a finite half-chain of size $N=50$ and various $\mu$. The lines show the $N \to \infty$ result
in \eqref{EHstag}.}
\label{fig:epsl}
\end{figure}

Finally, we discuss the spectrum of $C$, which is obtained completely analogously to the
previous section. The only difference is the extra factor $\kappa$ in the phase $\varphi_q$,
such that the EH and its spectrum read
\eq{
H = 4 \kappa K(\kappa') T, \qquad
\varepsilon_\ell =2\pi \frac{K(\kappa')}{K(\kappa)}
\begin{cases}
2\ell - \frac{1}{2} & \textrm{if $\mu>0$} \\ 
2\ell + \frac{1}{2} & \textrm{if $\mu<0$} \\ 
\end{cases}.
\label{EHstag}}
We have checked this result against numerical calculations for a finite open chain of total size $2N$.
The $\varepsilon_\ell$ are computed from the eigenvalues of the reduced correlation matrix of the half-chain
with size $N=50$, and shown in Fig.~\ref{fig:epsl} for various $\mu$. The comparison with the $N \to \infty$
result (lines) in \eqref{EHstag} shows an excellent agreement, which is expected for a gapped system with
$N \gg \xi$, i.e. when the size far exceeds the correlation length. The corresponding eigenvectors are very
similar to the dimerized case in Fig.~\ref{fig:psimdim}, and their normalization is analogous to \eqref{normpsimdim},
with the exchange $\delta' \to \kappa$ and an extra factor $\kappa$ coming from the size of the integration domain
on the real axis.

\section{Discussion\label{sec:disc}}

We have studied the EH in nonrelativistic free-fermion systems for a semi-infinite domain, which is described by the
deformed physical Hamiltonian with linearly increasing couplings. Its single-particle eigenstates have a remarkably
simple and universal structure, which holds both
for models defined in the continuum as well as on the lattice, and both for critical and gapped ground states.
Indeed, in all the cases considered, these are obtained as a superposition of the modes of the total system,
weighted with the inverse of their single-particle energy measured from the Fermi level.
Additionally, each mode receives an extra phase that is proportional to the integrated weight,
and the prefactor is related to the corresponding eigenvalue. The spectrum is continuous for critical systems,
and becomes discrete with an equidistant spacing for the gapped chains under study.
While for the dimerized chain this can also be obtained by duality with the transverse Ising chain 
and the corresponding CTM \cite{EDGTP20}, the result for the staggered chain is different and features a slightly shifted
level structure. In fact, it is unclear whether there exists a classical 2D integrable model whose CTM would
reproduce the same spectrum.

A further important observation is that, despite the nonrelativistic form of the models considered,
the BW form becomes exact, i.e. the EH is equal to the linearly deformed Hamiltonian up to a prefactor.
While in the critical case one finds the expected form \eqref{BW} of the BW theorem with $v$ given by the Fermi velocity,
for gapped systems the prefactor is nonuniversal. It is related to the behaviour of the inverse of the extra phase appearing
in the eigenfunctions, in particular its periodicity along the imaginary direction on the complex plane.
In contrast to the relativistic case, it carries a nontrivial dependence on the mass gap.

There are various possible extensions of this work. First, one could ask if the ansatz found for
a semi-infinite subsystem could be generalized to other geometries, such as the interval, where the entanglement
entropy has been studied thoroughly \cite{JK04,CMV11,MPST22}.
The commuting operator/matrix corresponds to the parabolic deformation of the Hamiltonian both in the continuum as well as
on the lattice \cite{EP13}, and their spectra and eigenfunctions were studied in detail \cite{Slepian65,Slepian78}. 
In turn, one finds that the EH is nontrivially related to the commuting operator.
While in the continuum the corrections vanish in the limit of a large interval \cite{Eisler24}, on the lattice the
structure of the EH is modified and involves longer range hopping \cite{EP17}, such that the CFT prediction
can be recovered only after an appropriate continuum limit \cite{ABCH17,ETP19}. It would be interesting to see,
if the eigenvectors of the EH could be cast in a form analogous to the one obtained here. Finding such a representation
could provide physical insight on the corrections observed in the EH beyond the CFT result. It could also help to
attack the massive case \cite{EDGTP20}, where no analytical QFT result is available so far and one has to rely on numerical
approaches \cite{BCM23}.

One could also address more general lattice models, e.g. when the dimerized hopping and staggered fields are
simultaneously present. A non-Hermitian version of such an SSH model has recently been studied in the context
of the EH, finding some nontrivial behaviour at the critical point \cite{RFC24}. Another interesting scenario where
commuting operators play a crucial role is the one of inhomogeneous hopping chains related to orthogonal
polynomials \cite{CNV19,CNV20,BCNPV24}. In the continuum limit, such inhomogeneous models can be
described by a CFT in a curved background metric \cite{DSVC17}, and the EH can be derived by appropriate conformal
mappings \cite{TRLS18,RSC22}, with a BW form shown to emerge in particular cases \cite{BE24}.
Generalizing our ansatz to the eigenvectors of the EH in these models could shed light on novel features.

Finally, it remains to understand how the ansatz could be adapted to interacting integrable systems, such as the XXZ
chain, where the EH in the gapped phase can be obtained via the CTM method and is known to have the BW form \cite{PKL99}.
The integrable generalization of the ansatz is further motivated by the fact, that the substitution applied in \eqref{qz}
is reminiscent of the rapidity parametrization. Finding the proper formulation could give a direct access to the EH
also in the critical phase, which could so far only be probed indirectly via numerical simulations \cite{DVZ18,GMSCD18,MSGDR19}.

\acknowledgments

We thank P-A. Bernard, R. Bonsignori, G. Parez, E. Tonni and L. Vinet for fruitful discussions and I. Peschel
for correspondence. The author acknowledges funding from the Austrian Science Fund (FWF) through
project No. P35434-N.

\newpage

\appendix 

\section{Calculation of commutators\label{app:comm}}

\subsection{Fermi gas}

We shall prove here that the differential operator \eqref{D} indeed commutes with the integral operator
$\hat K$ involving the sine kernel \eqref{Ksin} on the semi-infinite domain. One has
\eq{
-2\hat D \hat K f = (x\frac{\dd^2}{\dd x^2}+\frac{\dd}{\dd x}+q_F^2 x)\int_{0}^{\infty} dy \, K(x-y) f(y)=
\int_{0}^{\infty} dy \, f(y) (x\frac{\dd^2}{\dd y^2}-\frac{\dd}{\dd y}+q_F^2 x) K(x-y)
}
as well as
\eq{
-2\hat K \hat D f = \int_{0}^{\infty} dy \, K(x-y) (y\frac{\dd^2}{\dd y^2}+\frac{\dd}{\dd y}+q_F^2 y) f(y)=
\int_{0}^{\infty} dy \, f(y) (\frac{\dd^2}{\dd y^2} y-\frac{\dd}{\dd y}+q_F^2 y) K(x-y),
}
where in the last step we have integrated by parts. Note that there is a boundary contribution
$-K(x)f(0)$ from the first order derivative. This, however, is cancelled by the boundary term of
the second order derivative at the second partial integration, which gives
$\left.-f(y)\frac{\dd}{\dd y}[yK(x-y)]\right|_0^{\infty}=K(x)f(0)$. Note that one also needs to require
that the function $f(y)\to 0$ and its derivative $f'(y)\to 0$ as $y\to \infty$. One has further
\eq{
\frac{\dd^2}{\dd y^2} y K(x-y)=-2K'(x-y)+yK''(x-y),
}
such that the commutator reads
\eq{
-2[\hat D, \hat K] f=\int_{0}^{\infty} dy \, f(y) \left[
(x-y) K''(x-y) + 2K'(x-y)+ q_F^2 (x-y) K(x-y) \right].
\label{commfg}}
It is then easy to see that
\eq{
\begin{split}
&K'(x-y) = q_F\frac{\cos q_F(x-y)}{\pi (x-y)} - \frac{\sin q_F(x-y)}{\pi (x-y)^2}, \\
&K''(x-y) = -q_F^2\frac{\sin q_F(x-y)}{\pi (x-y)} - 2q_F\frac{\cos q_F(x-y)}{\pi (x-y)^2}+2\frac{\sin q_F(x-y)}{\pi (x-y)^3}
\end{split}}
and thus substituting into \eqref{commfg} the commutator vanishes.

\subsection{Hopping chain}

Next we show, that the tridiagonal matrix \eqref{T} with elements given by \eqref{Thom}
commutes with the discrete sine kernel in \eqref{C}. One has
\eq{
\begin{split}
&-2(TC)_{m,n}=
m \, C_{m+1,n} + (m-1) \, C_{m-1,n} - 2 \cos q_F (m-1/2) C_{m,n}
\\
&-2(CT)_{m,n}=
n \, C_{m,n+1} + (n-1) \, C_{m,n-1}- 2 \cos q_F (n-1/2) C_{m,n}
\end{split}
\label{TCCT}}
One can then introduce $r=m-n$ and rewrite
\eq{
-2[T,C]_{m,n}=
r (C_{m+1,n} + C_{m-1,n}) + C_{m+1,n}- C_{m-1,n} - 2 r \cos q_F C_{m,n}
\label{TCcomm}}
Furthermore one has
\eq{
\begin{split}
&C_{m+1,n} + C_{m-1,n} = 
\frac{\sin q_F(r+1)}{\pi (r+1)} + \frac{\sin q_F(r-1)}{\pi (r-1)}=
\frac{2r \sin q_F r \cos q_F - 2 \sin q_F \cos q_F r}{\pi (r^2-1)},\\
&C_{m+1,n} - C_{m-1,n} = 
\frac{\sin q_F(r+1)}{\pi (r+1)} - \frac{\sin q_F(r-1)}{\pi (r-1)}=
\frac{2r \sin q_F \cos q_F r - 2 \sin q_F r \cos q_F}{\pi (r^2-1)},
\end{split}}
and substituting into \eqref{TCcomm} one can easily recover $[T,C]=0$.

\subsection{Dimerized chain}

For the dimerized chain $T$ and $C$ have an alternating structure. However,
they both couple only sites over odd distances, such that their product is nonvanishing
only for even index distances. Assuming even indices one has
\eq{
\begin{split}
&-2(TC)_{2m,2n}=
(1-\delta) \, 2m \, C_{2m+1,2n} + (1+\delta) (2m-1) \, C_{2m-1,2n}
\\
&-2(CT)_{2m,2n}=
(1-\delta)\, 2n \, C_{2m,2n+1} + (1+\delta)(2n-1) \, C_{2m,2n-1}
\end{split}}
Furthermore, the correlation matrix elements have the structure
\eq{
C_{2m-1,2n} = \mathcal{C}_{r}+\delta \, \mathcal{S}_r, \qquad
C_{2m,2n+1} = \mathcal{C}_{r}-\delta \, \mathcal{S}_r,
}
where $r=2n+1-2m$ and we defined the integrals
\eq{
\mathcal{C}_r = \int_{-\pi/2}^{\pi/2} \frac{d q}{2\pi}
\frac{\cos qr \cos q}{\sqrt{\cos^2 q+\delta^2 \sin^2 q}} \, , \qquad
\mathcal{S}_r = \int_{-\pi/2}^{\pi/2} \frac{d q}{2\pi}
\frac{\sin qr \sin q}{\sqrt{\cos^2 q+\delta^2 \sin^2 q}} \, .
\label{crsr}}
Inserting these expressions, one obtains for the commutator matrix element
\begin{align}
-2[C,T]_{2m,2n} &=
(r-1)[\mathcal{C}_{r} + \mathcal{C}_{r-2} + \delta^2 (\mathcal{S}_{r} - \mathcal{S}_{r-2})]
+ (\mathcal{C}_{r} - \mathcal{C}_{r-2}) + \delta^2(\mathcal{S}_{r} + \mathcal{S}_{r-2}) \nonumber \\
&-(2m+2n-1) \delta (\mathcal{C}_{r}-\mathcal{C}_{r-2} + \mathcal{S}_{r}+\mathcal{S}_{r-2})
\label{commdim}
\end{align}
The first of these terms can be evaluated using trigonometric identities as
\begin{align}
(r-1)[\mathcal{C}_{r} + \mathcal{C}_{r-2} + \delta^2 (\mathcal{S}_{r} - \mathcal{S}_{r-2})]
&=\int_{-\pi/2}^{\pi/2} \frac{d q}{2\pi} (r-1) \cos q(r-1) \sqrt{\cos^2 q+\delta^2 \sin^2 q} \nonumber \\
&=\int_{-\pi/2}^{\pi/2} \frac{d q}{2\pi} (1-\delta^2 )\frac{\sin q(r-1) \sin 2q}{\sqrt{\cos^2 q+\delta^2 \sin^2 q}},
\end{align}
where in the second line we have integrated by parts. For the rest of the terms we need
the expressions
\eq{
\mathcal{S}_{r} + \mathcal{S}_{r-2} = -(\mathcal{C}_{r} - \mathcal{C}_{r-2})=
\int_{-\pi/2}^{\pi/2} \frac{d q}{2\pi} \frac{\sin q(r-1) \sin 2q}{\sqrt{\cos^2 q+\delta^2 \sin^2 q}}.
}
One can then immediately verify, that the first term in \eqref{commdim} cancels with the second
and third one in the first line, while the second line also gives zero. The proof for the case
of odd indices follows similarly.

\subsection{Staggered potential}

For the staggered potential the elements of the correlation matrix with $r=m-n$ read
\eq{
C_{m,n} - \frac{1}{2}\delta_{m,n} =
\begin{cases}
- (-1)^m \mu \int_{-\pi/2}^{\pi/2} \frac{d q}{2\pi}
\frac{\cos qr}{\sqrt{\cos^2 q+\mu^2}} & \textrm{$m-n$ even} \\
\int_{-\pi/2}^{\pi/2} \frac{d q}{2\pi}
\frac{\cos qr \cos q}{\sqrt{\cos^2 q+\mu^2}} & \textrm{$m-n$ odd} 
\end{cases}
\label{Cstag}}
The matrix products with the $T$ have a similar form as for the homogeneous chain
in \eqref{TCCT}, except for the alternation of the potential. However, when calculating
the commutator, one has now the property $C_{m-1,n}= \pm C_{m,n+1}$, where the
plus/minus sign corresponds to $m-n$ being even or odd. This yields
\eq{
-2[T,C]_{m,n}= 
\begin{cases}
(m-n)(C_{m+1,n} + C_{m-1,n} - (-1)^m 2\mu C_{m,n}) + C_{m+1,n}- C_{m-1,n} & \textrm{$m-n$ even} \\
(m+n-1) (C_{m+1,n} + C_{m-1,n} - (-1)^m 2\mu C_{m,n}) & \textrm{$m-n$ odd}
\end{cases}
\label{commstag}}
Considering the case $m-n$ even first, one finds from \eqref{Cstag}
\begin{align}
&r(C_{m+1,n} + C_{m-1,n} - (-1)^m 2\mu C_{m,n})=
\int_{-\pi/2}^{\pi/2} \frac{d q}{2\pi}
2r \cos qr \sqrt{\cos^2 q+\mu^2}, \label{Cstag1} \\
&C_{m+1,n}- C_{m-1,n} =-
\int_{-\pi/2}^{\pi/2} \frac{d q}{2\pi} \frac{\sin qr \sin 2q}{\sqrt{\cos^2 q+\mu^2}}.
\label{Cstag2}
\end{align}
Integrating by parts in \eqref{Cstag1}, one finds exactly the negative of \eqref{Cstag2},
and thus the commutator \eqref{commstag} vanishes. It is easy to check using \eqref{Cstag}
that the same holds true for $m-n$ odd as well.

\section{Relation to orthogonal polynomials\label{app:MP}}

Here we show that the eigenvectors \eqref{psim} are related to the so-called Meixner-Pollaczek
polynomials \cite{Koekoek10}, and derive the proportionality factor between them.
These hypergeometric orthogonal polynomials can be expressed as
\eq{
P^{(\alpha)}_{n}(x;\phi) = \frac{(2\alpha)_n}{n!} \ee^{in\phi} \,
_2F_1\Big(
\begin{array}{c}
-n, \,\, \lambda + ix \\
2\lambda
\end{array} ; 
1-\ee^{-2i\phi}\Big)
}
where $(2\alpha)_n$ is the Pochhammer symbol, and the first few polynomials are given by
\begin{align}
&P^{(\alpha)}_{0}(x;\phi) = 1 \nonumber \\
&P^{(\alpha)}_{1}(x;\phi) = 2(\alpha \cos \phi + x \sin \phi) \label{P012}\\
&P^{(\alpha)}_{2}(x;\phi) = x^2+\alpha^2 + (\alpha^2+\alpha-x^2)\cos(2\phi) + (1+2\lambda) x\sin (2\phi) \nonumber
\end{align}
The Meixner-Pollaczek polynomials satisfy the orthogonality relation with respect to a weight function
\eq{
\int_{-\infty}^{\infty} \dd x \, w(x;\alpha,\phi) \, 
P^{(\alpha)}_{n}(x;\phi)P^{(\alpha)}_{m}(x;\phi)=
\frac{2\pi \Gamma(n+2\alpha)}{(2\sin \phi)^{2\alpha}n!}\delta_{m,n},
\qquad
w(x;\alpha,\phi) = |\Gamma(\alpha+ix)|^2 \ee^{(2\phi-\pi)x}.
\label{orthMP}}

As pointed out in the main text, our eigenvectors correspond to $\alpha=1/2$, $\phi=q_F$, and $x=-\lambda/\sin q_F$,
there is, however, a proportionality factor to be fixed. A direct numerical comparison of the first few components
of $\psi_\lambda(m)$ and \eqref{P012} for various values of $\lambda$ and $q_F$ suggests the relation
\eq{
P^{(1/2)}_{n}(x;q_F) =
\sqrt{2} \ee^{\frac{q_F \lambda}{\sin q_F}}\cosh(\frac{\pi \lambda}{\sin q_F}) \psi_\lambda(n+1) \, .
}
We will now prove that the orthogonality relation \eqref{orthMP} is indeed satisfied with this choice.
Plugging in one has
\eq{
\pi\int_{-\infty}^{\infty} \dd x \,
(1+\ee^{-2\pi x})
\int_{-\infty}^{\infty} \frac{\dd z}{\pi} \, \frac{\ee^{iq(z)(m+1/2)}}{\sqrt{\cos q_F + \cosh (2z)}}
\int_{-\infty}^{\infty} \frac{\dd z'}{\pi} \, \frac{ \ee^{-iq(z')(n+1/2)}}{\sqrt{\cos q_F + \cosh (2z')}}
\ee^{i 2 x (z'-z)}.
}
The integral over $x$ can be carried out and reads
\eq{
\int_{-\infty}^{\infty} \dd x \,
(1+\ee^{-2\pi x}) \ee^{i 2 x (z'-z)}=
\pi [\delta(z'-z)+\delta(z'-z+i\pi)] \, .
\label{xint}}
The first delta function simply reproduces $\frac{\pi}{\sin q_F} C_{m,n}$. The second one requires an $i\pi$ shift between
the two variables. In order to understand its effect, let us shift the variables $z \to z + i \pi/2$
and $z' \to z' - i \pi/2$ to the upper and lower half-planes, respectively. The corresponding changes are
\eq{
\cosh (2z) \to -\cosh (2z), \qquad
\tanh (z) \to \coth (z), \qquad
}
and similarly for $z'$. In turn, the function $q(z)$ and its Jacobian transforms into
\eq{
\bar q(z) = 2 \atan[\tan(q_F/2)\coth (z)], \qquad
\frac{\dd \bar q}{\dd z} =\frac{2 \sin q_F}{\cos q_F - \cosh (2z)},
}
such that $\bar q(z)$ now maps the infinite line to the complement $\bar F$ of the Fermi sea.
The second delta function in \eqref{xint} thus yields the following contribution
\eq{
-\pi\int_{-\infty}^{\infty} \frac{\dd z}{2\pi} \, \frac{2}{\cos q_F - \cosh (2z)}
\ee^{i\bar q(z)(m-n)},
}
where the extra sign appears because of the branch cut of the square root.
The weight factor is thus proportional to the Jacobian of the transformation, 
and changing back to the $\bar q$ variable, one obtains the complementary sine kernel
\eq{
\frac{\pi}{\sin q_F} \bar C_{m,n} = \frac{\pi}{\sin q_F} (\delta_{m,n}- C_{m,n})
}
Adding the two pieces, we obtain the relation \eqref{orthMP}.

\newpage
\bibliography{ehsinf_refs}

\begin{thebibliography}{58}%
\makeatletter
\providecommand \@ifxundefined [1]{%
 \@ifx{#1\undefined}
}%
\providecommand \@ifnum [1]{%
 \ifnum #1\expandafter \@firstoftwo
 \else \expandafter \@secondoftwo
 \fi
}%
\providecommand \@ifx [1]{%
 \ifx #1\expandafter \@firstoftwo
 \else \expandafter \@secondoftwo
 \fi
}%
\providecommand \natexlab [1]{#1}%
\providecommand \enquote  [1]{``#1''}%
\providecommand \bibnamefont  [1]{#1}%
\providecommand \bibfnamefont [1]{#1}%
\providecommand \citenamefont [1]{#1}%
\providecommand \href@noop [0]{\@secondoftwo}%
\providecommand \href [0]{\begingroup \@sanitize@url \@href}%
\providecommand \@href[1]{\@@startlink{#1}\@@href}%
\providecommand \@@href[1]{\endgroup#1\@@endlink}%
\providecommand \@sanitize@url [0]{\catcode `\\12\catcode `\$12\catcode
  `\&12\catcode `\#12\catcode `\^12\catcode `\_12\catcode `\%12\relax}%
\providecommand \@@startlink[1]{}%
\providecommand \@@endlink[0]{}%
\providecommand \url  [0]{\begingroup\@sanitize@url \@url }%
\providecommand \@url [1]{\endgroup\@href {#1}{\urlprefix }}%
\providecommand \urlprefix  [0]{URL }%
\providecommand \Eprint [0]{\href }%
\providecommand \doibase [0]{https://doi.org/}%
\providecommand \selectlanguage [0]{\@gobble}%
\providecommand \bibinfo  [0]{\@secondoftwo}%
\providecommand \bibfield  [0]{\@secondoftwo}%
\providecommand \translation [1]{[#1]}%
\providecommand \BibitemOpen [0]{}%
\providecommand \bibitemStop [0]{}%
\providecommand \bibitemNoStop [0]{.\EOS\space}%
\providecommand \EOS [0]{\spacefactor3000\relax}%
\providecommand \BibitemShut  [1]{\csname bibitem#1\endcsname}%
\let\auto@bib@innerbib\@empty
\bibitem [{\citenamefont {Amico}\ \emph {et~al.}(2008)\citenamefont {Amico},
  \citenamefont {Fazio}, \citenamefont {Osterloh},\ and\ \citenamefont
  {Vedral}}]{AFOV08}%
  \BibitemOpen
  \bibfield  {author} {\bibinfo {author} {\bibfnamefont {L.}~\bibnamefont
  {Amico}}, \bibinfo {author} {\bibfnamefont {R.}~\bibnamefont {Fazio}},
  \bibinfo {author} {\bibfnamefont {A.}~\bibnamefont {Osterloh}},\ and\
  \bibinfo {author} {\bibfnamefont {V.}~\bibnamefont {Vedral}},\ }\bibfield
  {title} {\bibinfo {title} {Entanglement in many-body systems},\ }\href
  {https://doi.org/10.1103/RevModPhys.80.517} {\bibfield  {journal} {\bibinfo
  {journal} {Rev. Mod. Phys.}\ }\textbf {\bibinfo {volume} {80}},\ \bibinfo
  {pages} {517} (\bibinfo {year} {2008})}\BibitemShut {NoStop}%
\bibitem [{\citenamefont {Calabrese}\ \emph {et~al.}(2009)\citenamefont
  {Calabrese}, \citenamefont {Cardy},\ and\ \citenamefont {Doyon}}]{CCD09}%
  \BibitemOpen
  \bibfield  {author} {\bibinfo {author} {\bibfnamefont {P.}~\bibnamefont
  {Calabrese}}, \bibinfo {author} {\bibfnamefont {J.}~\bibnamefont {Cardy}},\
  and\ \bibinfo {author} {\bibfnamefont {B.}~\bibnamefont {Doyon}},\ }\bibfield
   {title} {\bibinfo {title} {Entanglement entropy in extended quantum
  systems},\ }\href {https://doi.org/10.1088/1751-8121/42/50/500301} {\bibfield
   {journal} {\bibinfo  {journal} {J. Phys. A: Math. Theor.}\ }\textbf
  {\bibinfo {volume} {42}},\ \bibinfo {pages} {500301} (\bibinfo {year}
  {2009})}\BibitemShut {NoStop}%
\bibitem [{\citenamefont {Eisert}\ \emph {et~al.}(2010)\citenamefont {Eisert},
  \citenamefont {Cramer},\ and\ \citenamefont {Plenio}}]{ECP10}%
  \BibitemOpen
  \bibfield  {author} {\bibinfo {author} {\bibfnamefont {J.}~\bibnamefont
  {Eisert}}, \bibinfo {author} {\bibfnamefont {M.}~\bibnamefont {Cramer}},\
  and\ \bibinfo {author} {\bibfnamefont {M.~B.}\ \bibnamefont {Plenio}},\
  }\bibfield  {title} {\bibinfo {title} {Colloquium: Area laws for the
  entanglement entropy},\ }\href {https://doi.org/10.1103/RevModPhys.82.277}
  {\bibfield  {journal} {\bibinfo  {journal} {Rev. Mod. Phys.}\ }\textbf
  {\bibinfo {volume} {82}},\ \bibinfo {pages} {277} (\bibinfo {year}
  {2010})}\BibitemShut {NoStop}%
\bibitem [{\citenamefont {Laflorencie}(2016)}]{Laflo16}%
  \BibitemOpen
  \bibfield  {author} {\bibinfo {author} {\bibfnamefont {N.}~\bibnamefont
  {Laflorencie}},\ }\bibfield  {title} {\bibinfo {title} {Quantum entanglement
  in condensed matter systems},\ }\href
  {https://doi.org/10.1016/j.physrep.2016.06.008} {\bibfield  {journal}
  {\bibinfo  {journal} {Phys. Rep.}\ }\textbf {\bibinfo {volume} {646}},\
  \bibinfo {pages} {1} (\bibinfo {year} {2016})}\BibitemShut {NoStop}%
\bibitem [{\citenamefont {Dalmonte}\ \emph {et~al.}(2022)\citenamefont
  {Dalmonte}, \citenamefont {Eisler}, \citenamefont {Falconi},\ and\
  \citenamefont {Vermersch}}]{DEFV22}%
  \BibitemOpen
  \bibfield  {author} {\bibinfo {author} {\bibfnamefont {M.}~\bibnamefont
  {Dalmonte}}, \bibinfo {author} {\bibfnamefont {V.}~\bibnamefont {Eisler}},
  \bibinfo {author} {\bibfnamefont {M.}~\bibnamefont {Falconi}},\ and\ \bibinfo
  {author} {\bibfnamefont {B.}~\bibnamefont {Vermersch}},\ }\bibfield  {title}
  {\bibinfo {title} {Entanglement {Hamiltonians}: From field theory to lattice
  models and experiments},\ }\href {https://doi.org/10.1002/andp.202200064}
  {\bibfield  {journal} {\bibinfo  {journal} {Ann. Phys.}\ }\textbf {\bibinfo
  {volume} {534}},\ \bibinfo {pages} {2200064} (\bibinfo {year}
  {2022})}\BibitemShut {NoStop}%
\bibitem [{\citenamefont {Kokail}\ \emph {et~al.}(2021)\citenamefont {Kokail},
  \citenamefont {van Bijnen}, \citenamefont {Elben}, \citenamefont
  {Vermersch},\ and\ \citenamefont {Zoller}}]{KBEVZ21}%
  \BibitemOpen
  \bibfield  {author} {\bibinfo {author} {\bibfnamefont {C.}~\bibnamefont
  {Kokail}}, \bibinfo {author} {\bibfnamefont {R.}~\bibnamefont {van Bijnen}},
  \bibinfo {author} {\bibfnamefont {A.}~\bibnamefont {Elben}}, \bibinfo
  {author} {\bibfnamefont {B.}~\bibnamefont {Vermersch}},\ and\ \bibinfo
  {author} {\bibfnamefont {P.}~\bibnamefont {Zoller}},\ }\bibfield  {title}
  {\bibinfo {title} {Entanglement {Hamiltonian} tomography in quantum
  simulation},\ }\href {https://doi.org/10.1038/s41567-021-01260-w} {\bibfield
  {journal} {\bibinfo  {journal} {Nat. Phys.}\ }\textbf {\bibinfo {volume}
  {17}},\ \bibinfo {pages} {936} (\bibinfo {year} {2021})}\BibitemShut
  {NoStop}%
\bibitem [{\citenamefont {Joshi}\ \emph {et~al.}(2023)\citenamefont {Joshi},
  \citenamefont {Kokail}, \citenamefont {van Bijnen}, \citenamefont {Kranzl},
  \citenamefont {Zache}, \citenamefont {Blatt}, \citenamefont {Roos},\ and\
  \citenamefont {Zoller}}]{Joshietal23}%
  \BibitemOpen
  \bibfield  {author} {\bibinfo {author} {\bibfnamefont {M.~K.}\ \bibnamefont
  {Joshi}}, \bibinfo {author} {\bibfnamefont {C.}~\bibnamefont {Kokail}},
  \bibinfo {author} {\bibfnamefont {R.}~\bibnamefont {van Bijnen}}, \bibinfo
  {author} {\bibfnamefont {F.}~\bibnamefont {Kranzl}}, \bibinfo {author}
  {\bibfnamefont {T.~V.}\ \bibnamefont {Zache}}, \bibinfo {author}
  {\bibfnamefont {R.}~\bibnamefont {Blatt}}, \bibinfo {author} {\bibfnamefont
  {C.~F.}\ \bibnamefont {Roos}},\ and\ \bibinfo {author} {\bibfnamefont
  {P.}~\bibnamefont {Zoller}},\ }\bibfield  {title} {\bibinfo {title}
  {Exploring large-scale entanglement in quantum simulation},\ }\href
  {https://doi.org/10.1038/s41586-023-06768-0} {\bibfield  {journal} {\bibinfo
  {journal} {Nature}\ }\textbf {\bibinfo {volume} {624}},\ \bibinfo {pages}
  {539} (\bibinfo {year} {2023})}\BibitemShut {NoStop}%
\bibitem [{\citenamefont {Haag}(1996)}]{Haag92}%
  \BibitemOpen
  \bibfield  {author} {\bibinfo {author} {\bibfnamefont {R.}~\bibnamefont
  {Haag}},\ }\href {https://doi.org/10.1007/978-3-642-61458-3} {\emph {\bibinfo
  {title} {Local quantum physics}}}\ (\bibinfo  {publisher} {Springer,
  Berlin},\ \bibinfo {year} {1996})\BibitemShut {NoStop}%
\bibitem [{\citenamefont {Borchers}(2000)}]{Borchers00}%
  \BibitemOpen
  \bibfield  {author} {\bibinfo {author} {\bibfnamefont {H.~J.}\ \bibnamefont
  {Borchers}},\ }\bibfield  {title} {\bibinfo {title} {{On revolutionizing
  quantum field theory with Tomita’s modular theory}},\ }\href
  {https://doi.org/10.1063/1.533323} {\bibfield  {journal} {\bibinfo  {journal}
  {J. Math. Phys.}\ }\textbf {\bibinfo {volume} {41}},\ \bibinfo {pages} {3604}
  (\bibinfo {year} {2000})}\BibitemShut {NoStop}%
\bibitem [{\citenamefont {Bisognano}\ and\ \citenamefont
  {Wichmann}(1975)}]{BW75}%
  \BibitemOpen
  \bibfield  {author} {\bibinfo {author} {\bibfnamefont {J.~J.}\ \bibnamefont
  {Bisognano}}\ and\ \bibinfo {author} {\bibfnamefont {E.~H.}\ \bibnamefont
  {Wichmann}},\ }\bibfield  {title} {\bibinfo {title} {On the duality condition
  for a {Hermitian} scalar field},\ }\href {https://doi.org/10.1063/1.522605}
  {\bibfield  {journal} {\bibinfo  {journal} {J. Math. Phys.}\ }\textbf
  {\bibinfo {volume} {16}},\ \bibinfo {pages} {985} (\bibinfo {year}
  {1975})}\BibitemShut {NoStop}%
\bibitem [{\citenamefont {Bisognano}\ and\ \citenamefont
  {Wichmann}(1976)}]{BW76}%
  \BibitemOpen
  \bibfield  {author} {\bibinfo {author} {\bibfnamefont {J.~J.}\ \bibnamefont
  {Bisognano}}\ and\ \bibinfo {author} {\bibfnamefont {E.~H.}\ \bibnamefont
  {Wichmann}},\ }\bibfield  {title} {\bibinfo {title} {On the duality condition
  for quantum fields},\ }\href {https://doi.org/10.1063/1.522898} {\bibfield
  {journal} {\bibinfo  {journal} {J. Math. Phys.}\ }\textbf {\bibinfo {volume}
  {17}},\ \bibinfo {pages} {303} (\bibinfo {year} {1976})}\BibitemShut
  {NoStop}%
\bibitem [{\citenamefont {Hislop}\ and\ \citenamefont {Longo}(1982)}]{HL82}%
  \BibitemOpen
  \bibfield  {author} {\bibinfo {author} {\bibfnamefont {P.~D.}\ \bibnamefont
  {Hislop}}\ and\ \bibinfo {author} {\bibfnamefont {R.}~\bibnamefont {Longo}},\
  }\bibfield  {title} {\bibinfo {title} {Modular structure of the local
  algebras associated with the free massless scalar field theory},\ }\href
  {https://doi.org/10.1007/BF01208372} {\bibfield  {journal} {\bibinfo
  {journal} {Comm. Math. Phys.}\ }\textbf {\bibinfo {volume} {84}},\ \bibinfo
  {pages} {71} (\bibinfo {year} {1982})}\BibitemShut {NoStop}%
\bibitem [{\citenamefont {Casini}\ \emph {et~al.}(2011)\citenamefont {Casini},
  \citenamefont {Huerta},\ and\ \citenamefont {Myers}}]{CHM11}%
  \BibitemOpen
  \bibfield  {author} {\bibinfo {author} {\bibfnamefont {H.}~\bibnamefont
  {Casini}}, \bibinfo {author} {\bibfnamefont {M.}~\bibnamefont {Huerta}},\
  and\ \bibinfo {author} {\bibfnamefont {R.~C.}\ \bibnamefont {Myers}},\
  }\bibfield  {title} {\bibinfo {title} {Towards a derivation of holographic
  entanglement entropy},\ }\href {https://doi.org/10.1007/JHEP05(2011)036}
  {\bibfield  {journal} {\bibinfo  {journal} {J. High Energy Phys.}\ }\textbf
  {\bibinfo {volume} {2011}}\bibinfo  {number} { (5)},\ \bibinfo {pages}
  {36}}\BibitemShut {NoStop}%
\bibitem [{\citenamefont {Wong}\ \emph {et~al.}(2013)\citenamefont {Wong},
  \citenamefont {Klich}, \citenamefont {Zayas},\ and\ \citenamefont
  {Vaman}}]{WKZV13}%
  \BibitemOpen
\bibfield  {number} {  }\bibfield  {author} {\bibinfo {author} {\bibfnamefont
  {G.}~\bibnamefont {Wong}}, \bibinfo {author} {\bibfnamefont {I.}~\bibnamefont
  {Klich}}, \bibinfo {author} {\bibfnamefont {L.~A.~P.}\ \bibnamefont
  {Zayas}},\ and\ \bibinfo {author} {\bibfnamefont {D.}~\bibnamefont {Vaman}},\
  }\bibfield  {title} {\bibinfo {title} {Entanglement temperature and
  entanglement entropy of excited states},\ }\href
  {https://doi.org/10.1007/JHEP12(2013)020} {\bibfield  {journal} {\bibinfo
  {journal} {J. High Energy Phys.}\ }\textbf {\bibinfo {volume} {2013}}\bibinfo
   {number} { (12)},\ \bibinfo {pages} {20}}\BibitemShut {NoStop}%
\bibitem [{\citenamefont {Cardy}\ and\ \citenamefont {Tonni}(2016)}]{CT16}%
  \BibitemOpen
\bibfield  {number} {  }\bibfield  {author} {\bibinfo {author} {\bibfnamefont
  {J.}~\bibnamefont {Cardy}}\ and\ \bibinfo {author} {\bibfnamefont
  {E.}~\bibnamefont {Tonni}},\ }\bibfield  {title} {\bibinfo {title}
  {Entanglement {Hamiltonians} in two-dimensional conformal field theory},\
  }\href {https://doi.org/10.1088/1742-5468/2016/12/123103} {\bibfield
  {journal} {\bibinfo  {journal} {J. Stat. Mech.: Theory Exp.}\ }\textbf
  {\bibinfo {volume} {2016}}\bibinfo  {number} { (12)},\ \bibinfo {pages}
  {123103}}\BibitemShut {NoStop}%
\bibitem [{\citenamefont {Peschel}\ \emph {et~al.}(1999)\citenamefont
  {Peschel}, \citenamefont {Kaulke},\ and\ \citenamefont {Legeza}}]{PKL99}%
  \BibitemOpen
\bibfield  {number} {  }\bibfield  {author} {\bibinfo {author} {\bibfnamefont
  {I.}~\bibnamefont {Peschel}}, \bibinfo {author} {\bibfnamefont
  {M.}~\bibnamefont {Kaulke}},\ and\ \bibinfo {author} {\bibfnamefont
  {{\"O}.}~\bibnamefont {Legeza}},\ }\bibfield  {title} {\bibinfo {title}
  {Density-matrix spectra for integrable models},\ }\href
  {https://doi.org/10.1002/andp.19995110203} {\bibfield  {journal} {\bibinfo
  {journal} {Ann. Phys. (Leipzig)}\ }\textbf {\bibinfo {volume} {8}},\ \bibinfo
  {pages} {153} (\bibinfo {year} {1999})}\BibitemShut {NoStop}%
\bibitem [{\citenamefont {Baxter}(1976)}]{Baxter76}%
  \BibitemOpen
  \bibfield  {author} {\bibinfo {author} {\bibfnamefont {R.~J.}\ \bibnamefont
  {Baxter}},\ }\bibfield  {title} {\bibinfo {title} {{Corner transfer matrices
  of the eight-vertex model: I. Low-temperature expansions and conjectured
  properties}},\ }\href {https://doi.org/10.1007/BF01020802} {\bibfield
  {journal} {\bibinfo  {journal} {J. Stat. Phys.}\ }\textbf {\bibinfo {volume}
  {15}},\ \bibinfo {pages} {485} (\bibinfo {year} {1976})}\BibitemShut
  {NoStop}%
\bibitem [{\citenamefont {Baxter}(1977)}]{Baxter77}%
  \BibitemOpen
  \bibfield  {author} {\bibinfo {author} {\bibfnamefont {R.~J.}\ \bibnamefont
  {Baxter}},\ }\bibfield  {title} {\bibinfo {title} {{Corner transfer matrices
  of the eight-vertex model: II.The Ising case}},\ }\href
  {https://doi.org/10.1007/BF01089373} {\bibfield  {journal} {\bibinfo
  {journal} {J. Stat. Phys.}\ }\textbf {\bibinfo {volume} {17}},\ \bibinfo
  {pages} {1} (\bibinfo {year} {1977})}\BibitemShut {NoStop}%
\bibitem [{\citenamefont {Baxter}(1982)}]{Baxter82}%
  \BibitemOpen
  \bibfield  {author} {\bibinfo {author} {\bibfnamefont {R.~J.}\ \bibnamefont
  {Baxter}},\ }\href@noop {} {\emph {\bibinfo {title} {Exactly Solved Models in
  Statistical Mechanics}}}\ (\bibinfo  {publisher} {Academic Press},\ \bibinfo
  {address} {London},\ \bibinfo {year} {1982})\BibitemShut {NoStop}%
\bibitem [{\citenamefont {Tetel'man}(1982)}]{Tetelman82}%
  \BibitemOpen
  \bibfield  {author} {\bibinfo {author} {\bibfnamefont {M.~G.}\ \bibnamefont
  {Tetel'man}},\ }\bibfield  {title} {\bibinfo {title} {Lorentz group for
  two-dimensional integrable lattice systems},\ }\href
  {http://www.jetp.ras.ru/cgi-bin/dn/e_055_02_0306.pdf} {\bibfield  {journal}
  {\bibinfo  {journal} {Sov. Phys. JETP}\ }\textbf {\bibinfo {volume} {55}},\
  \bibinfo {pages} {306} (\bibinfo {year} {1982})}\BibitemShut {NoStop}%
\bibitem [{\citenamefont {Thacker}(1986)}]{Thacker86}%
  \BibitemOpen
  \bibfield  {author} {\bibinfo {author} {\bibfnamefont {H.~B.}\ \bibnamefont
  {Thacker}},\ }\bibfield  {title} {\bibinfo {title} {Corner transfer matrices
  and lorentz invariance on a lattice},\ }\href
  {https://doi.org/10.1016/0167-2789(86)90196-X} {\bibfield  {journal}
  {\bibinfo  {journal} {Physica D}\ }\textbf {\bibinfo {volume} {18}},\
  \bibinfo {pages} {348} (\bibinfo {year} {1986})}\BibitemShut {NoStop}%
\bibitem [{\citenamefont {Itoyama}\ and\ \citenamefont {Thacker}(1987)}]{IT87}%
  \BibitemOpen
  \bibfield  {author} {\bibinfo {author} {\bibfnamefont {H.}~\bibnamefont
  {Itoyama}}\ and\ \bibinfo {author} {\bibfnamefont {H.~B.}\ \bibnamefont
  {Thacker}},\ }\bibfield  {title} {\bibinfo {title} {{Lattice Virasoro algebra
  and corner transfer matrices in the Baxter eight-vertex model}},\ }\href
  {https://doi.org/10.1103/PhysRevLett.58.1395} {\bibfield  {journal} {\bibinfo
   {journal} {Phys. Rev. Lett.}\ }\textbf {\bibinfo {volume} {58}},\ \bibinfo
  {pages} {1395} (\bibinfo {year} {1987})}\BibitemShut {NoStop}%
\bibitem [{\citenamefont {Thacker}\ and\ \citenamefont {Itoyama}(1988)}]{TI88}%
  \BibitemOpen
  \bibfield  {author} {\bibinfo {author} {\bibfnamefont {H.}~\bibnamefont
  {Thacker}}\ and\ \bibinfo {author} {\bibfnamefont {H.}~\bibnamefont
  {Itoyama}},\ }\bibfield  {title} {\bibinfo {title} {Integrability, conformal
  symmetry, and noncritical {Virasoro} algebras},\ }\href
  {https://doi.org/10.1016/0920-5632(88)90004-7} {\bibfield  {journal}
  {\bibinfo  {journal} {Nucl. Phys. B Proc. Suppl.}\ }\textbf {\bibinfo
  {volume} {5}},\ \bibinfo {pages} {9} (\bibinfo {year} {1988})}\BibitemShut
  {NoStop}%
\bibitem [{\citenamefont {Davies}(1988)}]{Davies88}%
  \BibitemOpen
  \bibfield  {author} {\bibinfo {author} {\bibfnamefont {B.}~\bibnamefont
  {Davies}},\ }\bibfield  {title} {\bibinfo {title} {Corner transfer matrices
  for the {Ising} model},\ }\href
  {https://doi.org/10.1016/0378-4371(88)90178-1} {\bibfield  {journal}
  {\bibinfo  {journal} {Physica A}\ }\textbf {\bibinfo {volume} {154}},\
  \bibinfo {pages} {1} (\bibinfo {year} {1988})}\BibitemShut {NoStop}%
\bibitem [{\citenamefont {Truong}\ and\ \citenamefont {Peschel}(1988)}]{TP88}%
  \BibitemOpen
  \bibfield  {author} {\bibinfo {author} {\bibfnamefont {T.~T.}\ \bibnamefont
  {Truong}}\ and\ \bibinfo {author} {\bibfnamefont {I.}~\bibnamefont
  {Peschel}},\ }\bibfield  {title} {\bibinfo {title} {Diagonalisation of corner
  transfer matrix by orthogonal polynomials},\ }\href
  {https://doi.org/10.1088/0305-4470/21/21/006} {\bibfield  {journal} {\bibinfo
   {journal} {J. Phys. A: Math. Gen.}\ }\textbf {\bibinfo {volume} {21}},\
  \bibinfo {pages} {L1029} (\bibinfo {year} {1988})}\BibitemShut {NoStop}%
\bibitem [{\citenamefont {Truong}\ and\ \citenamefont {Peschel}(1989)}]{TP89}%
  \BibitemOpen
  \bibfield  {author} {\bibinfo {author} {\bibfnamefont {T.~T.}\ \bibnamefont
  {Truong}}\ and\ \bibinfo {author} {\bibfnamefont {I.}~\bibnamefont
  {Peschel}},\ }\bibfield  {title} {\bibinfo {title} {Diagonalisation of
  finite-size corner transfer matrices and related spin chains},\ }\href
  {https://doi.org/10.1007/BF01313574} {\bibfield  {journal} {\bibinfo
  {journal} {Z. Phys. B}\ }\textbf {\bibinfo {volume} {75}},\ \bibinfo {pages}
  {119} (\bibinfo {year} {1989})}\BibitemShut {NoStop}%
\bibitem [{\citenamefont {Truong}\ and\ \citenamefont {Peschel}(1990)}]{TP90}%
  \BibitemOpen
  \bibfield  {author} {\bibinfo {author} {\bibfnamefont {T.}~\bibnamefont
  {Truong}}\ and\ \bibinfo {author} {\bibfnamefont {I.}~\bibnamefont
  {Peschel}},\ }\bibfield  {title} {\bibinfo {title} {The corner transfer
  matrix of some free-fermion systems and {Meixner's} polynomials},\ }\href
  {https://doi.org/10.1142/S0217979290000413} {\bibfield  {journal} {\bibinfo
  {journal} {Int. J. Mod. Phys. B}\ }\textbf {\bibinfo {volume} {04}},\
  \bibinfo {pages} {895} (\bibinfo {year} {1990})}\BibitemShut {NoStop}%
\bibitem [{\citenamefont {Eckle}\ and\ \citenamefont {Truong}(1992)}]{ET92}%
  \BibitemOpen
  \bibfield  {author} {\bibinfo {author} {\bibfnamefont {H.~P.}\ \bibnamefont
  {Eckle}}\ and\ \bibinfo {author} {\bibfnamefont {T.~T.}\ \bibnamefont
  {Truong}},\ }\bibfield  {title} {\bibinfo {title} {Corner transfer matrix of
  a critical free fermion system},\ }\href
  {https://doi.org/10.1088/0305-4470/25/9/006} {\bibfield  {journal} {\bibinfo
  {journal} {J. Phys. A: Math. Gen.}\ }\textbf {\bibinfo {volume} {25}},\
  \bibinfo {pages} {L535} (\bibinfo {year} {1992})}\BibitemShut {NoStop}%
\bibitem [{\citenamefont {Eisler}\ and\ \citenamefont {Peschel}(2018)}]{EP18}%
  \BibitemOpen
  \bibfield  {author} {\bibinfo {author} {\bibfnamefont {V.}~\bibnamefont
  {Eisler}}\ and\ \bibinfo {author} {\bibfnamefont {I.}~\bibnamefont
  {Peschel}},\ }\bibfield  {title} {\bibinfo {title} {Properties of the
  entanglement {Hamiltonian} for finite free-fermion chains},\ }\href
  {https://doi.org/10.1088/1742-5468/aace2b} {\bibfield  {journal} {\bibinfo
  {journal} {J. Stat. Mech.: Theory Exp.}\ }\textbf {\bibinfo {volume}
  {2018}}\bibinfo  {number} { (10)},\ \bibinfo {pages} {104001}}\BibitemShut
  {NoStop}%
\bibitem [{\citenamefont {Peschel}(2003)}]{Peschel03}%
  \BibitemOpen
\bibfield  {number} {  }\bibfield  {author} {\bibinfo {author} {\bibfnamefont
  {I.}~\bibnamefont {Peschel}},\ }\bibfield  {title} {\bibinfo {title}
  {Calculation of reduced density matrices from correlation functions},\ }\href
  {https://doi.org/10.1088/0305-4470/36/14/101} {\bibfield  {journal} {\bibinfo
   {journal} {J. Phys. A: Math. Gen.}\ }\textbf {\bibinfo {volume} {36}},\
  \bibinfo {pages} {L205} (\bibinfo {year} {2003})}\BibitemShut {NoStop}%
\bibitem [{\citenamefont {Peschel}\ and\ \citenamefont {Eisler}(2009)}]{PE09}%
  \BibitemOpen
  \bibfield  {author} {\bibinfo {author} {\bibfnamefont {I.}~\bibnamefont
  {Peschel}}\ and\ \bibinfo {author} {\bibfnamefont {V.}~\bibnamefont
  {Eisler}},\ }\bibfield  {title} {\bibinfo {title} {Reduced density matrices
  and entanglement entropy in free lattice models},\ }\href
  {https://doi.org/10.1088/1751-8113/42/50/504003} {\bibfield  {journal}
  {\bibinfo  {journal} {J. Phys. A: Math. Theor.}\ }\textbf {\bibinfo {volume}
  {42}},\ \bibinfo {pages} {504003} (\bibinfo {year} {2009})}\BibitemShut
  {NoStop}%
\bibitem [{\citenamefont {Slepian}\ and\ \citenamefont {Pollak}(1961)}]{SP61}%
  \BibitemOpen
  \bibfield  {author} {\bibinfo {author} {\bibfnamefont {D.}~\bibnamefont
  {Slepian}}\ and\ \bibinfo {author} {\bibfnamefont {H.~O.}\ \bibnamefont
  {Pollak}},\ }\bibfield  {title} {\bibinfo {title} {Prolate spheroidal wave
  functions, {Fourier} analysis and uncertainty - {I}},\ }\href
  {https://doi.org/10.1002/j.1538-7305.1961.tb03976.x} {\bibfield  {journal}
  {\bibinfo  {journal} {Bell Syst. Tech. J.}\ }\textbf {\bibinfo {volume}
  {40}},\ \bibinfo {pages} {43} (\bibinfo {year} {1961})}\BibitemShut {NoStop}%
\bibitem [{\citenamefont {{Abramowitz}}\ and\ \citenamefont
  {{Stegun}}(1964)}]{AS64}%
  \BibitemOpen
  \bibfield  {author} {\bibinfo {author} {\bibfnamefont {M.}~\bibnamefont
  {{Abramowitz}}}\ and\ \bibinfo {author} {\bibfnamefont {I.~A.}\ \bibnamefont
  {{Stegun}}},\ }\href@noop {} {\emph {\bibinfo {title} {Handbook of
  Mathematical Functions with Formulas, Graphs, and Mathematical Tables}}},\
  \bibinfo {edition} {ninth dover printing, tenth gpo printing}\ ed.\ (\bibinfo
   {publisher} {Dover},\ \bibinfo {address} {New York City},\ \bibinfo {year}
  {1964})\BibitemShut {NoStop}%
\bibitem [{\citenamefont {Slepian}(1978)}]{Slepian78}%
  \BibitemOpen
  \bibfield  {author} {\bibinfo {author} {\bibfnamefont {D.}~\bibnamefont
  {Slepian}},\ }\bibfield  {title} {\bibinfo {title} {Prolate spheroidal wave
  functions, {Fourier} analysis, and uncertainty — {V}: The discrete case},\
  }\href {https://doi.org/10.1002/j.1538-7305.1978.tb02104.x} {\bibfield
  {journal} {\bibinfo  {journal} {Bell Syst. Tech. J.}\ }\textbf {\bibinfo
  {volume} {57}},\ \bibinfo {pages} {1371} (\bibinfo {year}
  {1978})}\BibitemShut {NoStop}%
\bibitem [{\citenamefont {Peschel}(2004)}]{Peschel04}%
  \BibitemOpen
  \bibfield  {author} {\bibinfo {author} {\bibfnamefont {I.}~\bibnamefont
  {Peschel}},\ }\bibfield  {title} {\bibinfo {title} {On the reduced density
  matrix for a chain of free electrons},\ }\href
  {https://doi.org/10.1088/1742-5468/2004/06/P06004} {\bibfield  {journal}
  {\bibinfo  {journal} {J. Stat. Mech.: Theory Exp.}\ }\textbf {\bibinfo
  {volume} {2004}}\bibinfo  {number} { (06)},\ \bibinfo {pages}
  {P06004}}\BibitemShut {NoStop}%
\bibitem [{\citenamefont {Koekoek}\ \emph {et~al.}(2010)\citenamefont
  {Koekoek}, \citenamefont {Lesky},\ and\ \citenamefont
  {Swarttouw}}]{Koekoek10}%
  \BibitemOpen
\bibfield  {number} {  }\bibfield  {author} {\bibinfo {author} {\bibfnamefont
  {R.}~\bibnamefont {Koekoek}}, \bibinfo {author} {\bibfnamefont {P.~A.}\
  \bibnamefont {Lesky}},\ and\ \bibinfo {author} {\bibfnamefont {R.~F.}\
  \bibnamefont {Swarttouw}},\ }\href
  {https://doi.org/10.1007/978-3-642-05014-5} {\emph {\bibinfo {title}
  {Hypergeometric orthogonal polynomials and their $q$-analogues}}}\ (\bibinfo
  {publisher} {Springer},\ \bibinfo {year} {2010})\BibitemShut {NoStop}%
\bibitem [{\citenamefont {Eisler}\ \emph {et~al.}(2020)\citenamefont {Eisler},
  \citenamefont {{Di Giulio}}, \citenamefont {Tonni},\ and\ \citenamefont
  {Peschel}}]{EDGTP20}%
  \BibitemOpen
  \bibfield  {author} {\bibinfo {author} {\bibfnamefont {V.}~\bibnamefont
  {Eisler}}, \bibinfo {author} {\bibfnamefont {G.}~\bibnamefont {{Di Giulio}}},
  \bibinfo {author} {\bibfnamefont {E.}~\bibnamefont {Tonni}},\ and\ \bibinfo
  {author} {\bibfnamefont {I.}~\bibnamefont {Peschel}},\ }\bibfield  {title}
  {\bibinfo {title} {Entanglement {Hamiltonians} for non-critical quantum
  chains},\ }\href {https://doi.org/10.1088/1742-5468/abb4da} {\bibinfo
  {journal} {J. Stat. Mech.: Theory Exp.}\ ,\ \bibinfo {eid}
  {103102}}\BibitemShut {NoStop}%
\bibitem [{\citenamefont {Jin}\ and\ \citenamefont {Korepin}(2004)}]{JK04}%
  \BibitemOpen
\bibfield  {journal} {  }\bibfield  {author} {\bibinfo {author} {\bibfnamefont
  {B.~Q.}\ \bibnamefont {Jin}}\ and\ \bibinfo {author} {\bibfnamefont {V.~E.}\
  \bibnamefont {Korepin}},\ }\bibfield  {title} {\bibinfo {title} {Quantum spin
  chain, {Toeplitz} determinants and the {Fisher---Hartwig} conjecture},\
  }\href {https://doi.org/10.1023/B:JOSS.0000037230.37166.42} {\bibfield
  {journal} {\bibinfo  {journal} {J. Stat. Phys.}\ }\textbf {\bibinfo {volume}
  {116}},\ \bibinfo {pages} {79} (\bibinfo {year} {2004})}\BibitemShut
  {NoStop}%
\bibitem [{\citenamefont {Calabrese}\ \emph {et~al.}(2011)\citenamefont
  {Calabrese}, \citenamefont {Mintchev},\ and\ \citenamefont {Vicari}}]{CMV11}%
  \BibitemOpen
  \bibfield  {author} {\bibinfo {author} {\bibfnamefont {P.}~\bibnamefont
  {Calabrese}}, \bibinfo {author} {\bibfnamefont {M.}~\bibnamefont
  {Mintchev}},\ and\ \bibinfo {author} {\bibfnamefont {E.}~\bibnamefont
  {Vicari}},\ }\bibfield  {title} {\bibinfo {title} {Entanglement entropy of
  one-dimensional gases},\ }\href
  {https://doi.org/10.1103/PhysRevLett.107.020601} {\bibfield  {journal}
  {\bibinfo  {journal} {Phys. Rev. Lett.}\ }\textbf {\bibinfo {volume} {107}},\
  \bibinfo {pages} {020601} (\bibinfo {year} {2011})}\BibitemShut {NoStop}%
\bibitem [{\citenamefont {Mintchev}\ \emph {et~al.}(2022)\citenamefont
  {Mintchev}, \citenamefont {Pontello}, \citenamefont {Sartori},\ and\
  \citenamefont {Tonni}}]{MPST22}%
  \BibitemOpen
  \bibfield  {author} {\bibinfo {author} {\bibfnamefont {M.}~\bibnamefont
  {Mintchev}}, \bibinfo {author} {\bibfnamefont {D.}~\bibnamefont {Pontello}},
  \bibinfo {author} {\bibfnamefont {A.}~\bibnamefont {Sartori}},\ and\ \bibinfo
  {author} {\bibfnamefont {E.}~\bibnamefont {Tonni}},\ }\bibfield  {title}
  {\bibinfo {title} {Entanglement entropies of an interval in the free
  {Schr{\"o}dinger} field theory at finite density},\ }\href
  {https://doi.org/10.1007/JHEP07(2022)120} {\bibfield  {journal} {\bibinfo
  {journal} {J. High Energy Phys.}\ }\textbf {\bibinfo {volume} {2022}}\bibinfo
   {number} { (7)},\ \bibinfo {pages} {120}}\BibitemShut {NoStop}%
\bibitem [{\citenamefont {Eisler}\ and\ \citenamefont {Peschel}(2013)}]{EP13}%
  \BibitemOpen
\bibfield  {number} {  }\bibfield  {author} {\bibinfo {author} {\bibfnamefont
  {V.}~\bibnamefont {Eisler}}\ and\ \bibinfo {author} {\bibfnamefont
  {I.}~\bibnamefont {Peschel}},\ }\bibfield  {title} {\bibinfo {title}
  {Free-fermion entanglement and spheroidal functions},\ }\href
  {https://doi.org/10.1088/1742-5468/2013/04/P04028} {\bibfield  {journal}
  {\bibinfo  {journal} {J. Stat. Mech.: Theory Exp.}\ }\textbf {\bibinfo
  {volume} {2013}}\bibinfo  {number} { (04)},\ \bibinfo {pages}
  {P04028}}\BibitemShut {NoStop}%
\bibitem [{\citenamefont {Slepian}(1965)}]{Slepian65}%
  \BibitemOpen
\bibfield  {number} {  }\bibfield  {author} {\bibinfo {author} {\bibfnamefont
  {D.}~\bibnamefont {Slepian}},\ }\bibfield  {title} {\bibinfo {title} {Some
  asymptotic expansions for prolate spheroidal wave functions},\ }\href
  {https://doi.org/10.1002/sapm196544199} {\bibfield  {journal} {\bibinfo
  {journal} {J. Math. and Phys.}\ }\textbf {\bibinfo {volume} {44}},\ \bibinfo
  {pages} {99} (\bibinfo {year} {1965})}\BibitemShut {NoStop}%
\bibitem [{\citenamefont {Eisler}(2024)}]{Eisler24}%
  \BibitemOpen
  \bibfield  {author} {\bibinfo {author} {\bibfnamefont {V.}~\bibnamefont
  {Eisler}},\ }\bibfield  {title} {\bibinfo {title} {{Entanglement Hamiltonian
  of a nonrelativistic Fermi gas}},\ }\href
  {https://doi.org/10.1103/PhysRevB.109.L201113} {\bibfield  {journal}
  {\bibinfo  {journal} {Phys. Rev. B}\ }\textbf {\bibinfo {volume} {109}},\
  \bibinfo {pages} {L201113} (\bibinfo {year} {2024})}\BibitemShut {NoStop}%
\bibitem [{\citenamefont {Eisler}\ and\ \citenamefont {Peschel}(2017)}]{EP17}%
  \BibitemOpen
  \bibfield  {author} {\bibinfo {author} {\bibfnamefont {V.}~\bibnamefont
  {Eisler}}\ and\ \bibinfo {author} {\bibfnamefont {I.}~\bibnamefont
  {Peschel}},\ }\bibfield  {title} {\bibinfo {title} {Analytical results for
  the entanglement {Hamiltonian} of a free-fermion chain},\ }\href
  {https://doi.org/10.1088/1751-8121/aa76b5} {\bibfield  {journal} {\bibinfo
  {journal} {J. Phys. A: Math. Theor.}\ }\textbf {\bibinfo {volume} {50}},\
  \bibinfo {pages} {284003} (\bibinfo {year} {2017})}\BibitemShut {NoStop}%
\bibitem [{\citenamefont {Arias}\ \emph {et~al.}(2017)\citenamefont {Arias},
  \citenamefont {Blanco}, \citenamefont {Casini},\ and\ \citenamefont
  {Huerta}}]{ABCH17}%
  \BibitemOpen
  \bibfield  {author} {\bibinfo {author} {\bibfnamefont {R.~E.}\ \bibnamefont
  {Arias}}, \bibinfo {author} {\bibfnamefont {D.~D.}\ \bibnamefont {Blanco}},
  \bibinfo {author} {\bibfnamefont {H.}~\bibnamefont {Casini}},\ and\ \bibinfo
  {author} {\bibfnamefont {M.}~\bibnamefont {Huerta}},\ }\bibfield  {title}
  {\bibinfo {title} {Local temperatures and local terms in modular
  {Hamiltonians}},\ }\href {https://doi.org/10.1103/PhysRevD.95.065005}
  {\bibfield  {journal} {\bibinfo  {journal} {Phys. Rev. D}\ }\textbf {\bibinfo
  {volume} {95}},\ \bibinfo {pages} {065005} (\bibinfo {year}
  {2017})}\BibitemShut {NoStop}%
\bibitem [{\citenamefont {Eisler}\ \emph {et~al.}(2019)\citenamefont {Eisler},
  \citenamefont {Tonni},\ and\ \citenamefont {Peschel}}]{ETP19}%
  \BibitemOpen
  \bibfield  {author} {\bibinfo {author} {\bibfnamefont {V.}~\bibnamefont
  {Eisler}}, \bibinfo {author} {\bibfnamefont {E.}~\bibnamefont {Tonni}},\ and\
  \bibinfo {author} {\bibfnamefont {I.}~\bibnamefont {Peschel}},\ }\bibfield
  {title} {\bibinfo {title} {On the continuum limit of the entanglement
  {Hamiltonian}},\ }\href {https://doi.org/10.1088/1742-5468/ab1f0e} {\bibfield
   {journal} {\bibinfo  {journal} {J. Stat. Mech.: Theory Exp.}\ }\textbf
  {\bibinfo {volume} {2019}}\bibinfo  {number} { (7)},\ \bibinfo {pages}
  {073101}}\BibitemShut {NoStop}%
\bibitem [{\citenamefont {Bostelmann}\ \emph {et~al.}(2023)\citenamefont
  {Bostelmann}, \citenamefont {Cadamuro},\ and\ \citenamefont {Minz}}]{BCM23}%
  \BibitemOpen
\bibfield  {number} {  }\bibfield  {author} {\bibinfo {author} {\bibfnamefont
  {H.}~\bibnamefont {Bostelmann}}, \bibinfo {author} {\bibfnamefont
  {D.}~\bibnamefont {Cadamuro}},\ and\ \bibinfo {author} {\bibfnamefont
  {C.}~\bibnamefont {Minz}},\ }\bibfield  {title} {\bibinfo {title} {On the
  mass dependence of the modular operator for a double cone},\ }\href
  {https://doi.org/10.1007/s00023-023-01311-3} {\bibfield  {journal} {\bibinfo
  {journal} {Ann. Henri Poincar{\'e}}\ }\textbf {\bibinfo {volume} {24}},\
  \bibinfo {pages} {3031} (\bibinfo {year} {2023})}\BibitemShut {NoStop}%
\bibitem [{\citenamefont {Rottoli}\ \emph {et~al.}(2024)\citenamefont
  {Rottoli}, \citenamefont {Fossati},\ and\ \citenamefont {Calabrese}}]{RFC24}%
  \BibitemOpen
  \bibfield  {author} {\bibinfo {author} {\bibfnamefont {F.}~\bibnamefont
  {Rottoli}}, \bibinfo {author} {\bibfnamefont {M.}~\bibnamefont {Fossati}},\
  and\ \bibinfo {author} {\bibfnamefont {P.}~\bibnamefont {Calabrese}},\
  }\bibfield  {title} {\bibinfo {title} {{Entanglement Hamiltonian in the
  non-Hermitian SSH model}},\ }\href {https://doi.org/10.1088/1742-5468/ad4860}
  {\bibfield  {journal} {\bibinfo  {journal} {J. Stat. Mech.: Theory Exp.}\
  }\textbf {\bibinfo {volume} {2024}},\ \bibinfo {pages} {063102}}\BibitemShut
  {NoStop}%
\bibitem [{\citenamefont {Crampé}\ \emph {et~al.}(2019)\citenamefont
  {Crampé}, \citenamefont {Nepomechie},\ and\ \citenamefont {Vinet}}]{CNV19}%
  \BibitemOpen
  \bibfield  {author} {\bibinfo {author} {\bibfnamefont {N.}~\bibnamefont
  {Crampé}}, \bibinfo {author} {\bibfnamefont {R.~I.}\ \bibnamefont
  {Nepomechie}},\ and\ \bibinfo {author} {\bibfnamefont {L.}~\bibnamefont
  {Vinet}},\ }\bibfield  {title} {\bibinfo {title} {Free-fermion entanglement
  and orthogonal polynomials},\ }\href
  {https://doi.org/10.1088/1742-5468/ab3787} {\bibfield  {journal} {\bibinfo
  {journal} {J. Stat. Mech.: Theory Exp.}\ }\textbf {\bibinfo {volume}
  {2019}}\bibinfo  {number} { (9)},\ \bibinfo {pages} {093101}}\BibitemShut
  {NoStop}%
\bibitem [{\citenamefont {Cramp\'{e}}\ \emph {et~al.}(2021)\citenamefont
  {Cramp\'{e}}, \citenamefont {Nepomechie},\ and\ \citenamefont
  {Vinet}}]{CNV20}%
  \BibitemOpen
\bibfield  {number} {  }\bibfield  {author} {\bibinfo {author} {\bibfnamefont
  {N.}~\bibnamefont {Cramp\'{e}}}, \bibinfo {author} {\bibfnamefont {R.~I.}\
  \bibnamefont {Nepomechie}},\ and\ \bibinfo {author} {\bibfnamefont
  {L.}~\bibnamefont {Vinet}},\ }\bibfield  {title} {\bibinfo {title}
  {Entanglement in fermionic chains and bispectrality},\ }\href
  {https://doi.org/10.1142/S0129055X21400018} {\bibfield  {journal} {\bibinfo
  {journal} {Rev. Math. Phys.}\ }\textbf {\bibinfo {volume} {33}},\ \bibinfo
  {pages} {2140001} (\bibinfo {year} {2021})}\BibitemShut {NoStop}%
\bibitem [{\citenamefont {Bernard}\ \emph {et~al.}(2024)\citenamefont
  {Bernard}, \citenamefont {Crampé}, \citenamefont {Nepomechie}, \citenamefont
  {Parez},\ and\ \citenamefont {Vinet}}]{BCNPV24}%
  \BibitemOpen
  \bibfield  {author} {\bibinfo {author} {\bibfnamefont {P.-A.}\ \bibnamefont
  {Bernard}}, \bibinfo {author} {\bibfnamefont {N.}~\bibnamefont {Crampé}},
  \bibinfo {author} {\bibfnamefont {R.}~\bibnamefont {Nepomechie}}, \bibinfo
  {author} {\bibfnamefont {G.}~\bibnamefont {Parez}},\ and\ \bibinfo {author}
  {\bibfnamefont {L.}~\bibnamefont {Vinet}},\ }\href@noop {} {\bibinfo {title}
  {Entanglement of free-fermion systems, signal processing and algebraic
  combinatorics}} (\bibinfo {year} {2024}),\ \Eprint
  {https://arxiv.org/abs/2401.07150} {arXiv:2401.07150} \BibitemShut {NoStop}%
\bibitem [{\citenamefont {Dubail}\ \emph {et~al.}(2017)\citenamefont {Dubail},
  \citenamefont {\relax{J-M.} St\'ephan}, \citenamefont {Viti},\ and\
  \citenamefont {Calabrese}}]{DSVC17}%
  \BibitemOpen
  \bibfield  {author} {\bibinfo {author} {\bibfnamefont {J.}~\bibnamefont
  {Dubail}}, \bibinfo {author} {\bibnamefont {\relax{J-M.} St\'ephan}},
  \bibinfo {author} {\bibfnamefont {J.}~\bibnamefont {Viti}},\ and\ \bibinfo
  {author} {\bibfnamefont {P.}~\bibnamefont {Calabrese}},\ }\bibfield  {title}
  {\bibinfo {title} {Conformal field theory for inhomogeneous one-dimensional
  quantum systems: the example of non-interacting fermi gases},\ }\href
  {https://doi.org/10.21468/SciPostPhys.2.1.002} {\bibfield  {journal}
  {\bibinfo  {journal} {SciPost Phys.}\ }\textbf {\bibinfo {volume} {2}},\
  \bibinfo {pages} {002} (\bibinfo {year} {2017})}\BibitemShut {NoStop}%
\bibitem [{\citenamefont {Tonni}\ \emph {et~al.}(2018)\citenamefont {Tonni},
  \citenamefont {Rodríguez-Laguna},\ and\ \citenamefont {Sierra}}]{TRLS18}%
  \BibitemOpen
  \bibfield  {author} {\bibinfo {author} {\bibfnamefont {E.}~\bibnamefont
  {Tonni}}, \bibinfo {author} {\bibfnamefont {J.}~\bibnamefont
  {Rodríguez-Laguna}},\ and\ \bibinfo {author} {\bibfnamefont
  {G.}~\bibnamefont {Sierra}},\ }\bibfield  {title} {\bibinfo {title}
  {Entanglement hamiltonian and entanglement contour in inhomogeneous 1d
  critical systems},\ }\href {https://doi.org/10.1088/1742-5468/aab67d}
  {\bibfield  {journal} {\bibinfo  {journal} {J. Stat. Mech.: Theory Exp.}\
  }\textbf {\bibinfo {volume} {2018}}\bibinfo  {number} { (4)},\ \bibinfo
  {pages} {043105}}\BibitemShut {NoStop}%
\bibitem [{\citenamefont {Rottoli}\ \emph {et~al.}(2022)\citenamefont
  {Rottoli}, \citenamefont {Scopa},\ and\ \citenamefont {Calabrese}}]{RSC22}%
  \BibitemOpen
\bibfield  {number} {  }\bibfield  {author} {\bibinfo {author} {\bibfnamefont
  {F.}~\bibnamefont {Rottoli}}, \bibinfo {author} {\bibfnamefont
  {S.}~\bibnamefont {Scopa}},\ and\ \bibinfo {author} {\bibfnamefont
  {P.}~\bibnamefont {Calabrese}},\ }\bibfield  {title} {\bibinfo {title}
  {Entanglement {Hamiltonian} during a domain wall melting in the free {Fermi}
  chain},\ }\href {https://doi.org/10.1088/1742-5468/ac72a1} {\bibfield
  {journal} {\bibinfo  {journal} {J. Stat. Mech.: Theory Exp.}\ }\textbf
  {\bibinfo {volume} {2022}}\bibinfo  {number} { (6)},\ \bibinfo {pages}
  {063103}}\BibitemShut {NoStop}%
\bibitem [{\citenamefont {Bonsignori}\ and\ \citenamefont
  {Eisler}(2024)}]{BE24}%
  \BibitemOpen
\bibfield  {number} {  }\bibfield  {author} {\bibinfo {author} {\bibfnamefont
  {R.}~\bibnamefont {Bonsignori}}\ and\ \bibinfo {author} {\bibfnamefont
  {V.}~\bibnamefont {Eisler}},\ }\bibfield  {title} {\bibinfo {title}
  {Entanglement {Hamiltonian} for inhomogeneous free fermions},\ }\href
  {https://doi.org/10.1088/1751-8121/ad5501} {\bibfield  {journal} {\bibinfo
  {journal} {J. Phys. A: Math. Theor.}\ }\textbf {\bibinfo {volume} {57}},\
  \bibinfo {pages} {275001} (\bibinfo {year} {2024})}\BibitemShut {NoStop}%
\bibitem [{\citenamefont {Dalmonte}\ \emph {et~al.}(2018)\citenamefont
  {Dalmonte}, \citenamefont {Vermersch},\ and\ \citenamefont {Zoller}}]{DVZ18}%
  \BibitemOpen
  \bibfield  {author} {\bibinfo {author} {\bibfnamefont {M.}~\bibnamefont
  {Dalmonte}}, \bibinfo {author} {\bibfnamefont {B.}~\bibnamefont
  {Vermersch}},\ and\ \bibinfo {author} {\bibfnamefont {P.}~\bibnamefont
  {Zoller}},\ }\bibfield  {title} {\bibinfo {title} {Quantum simulation and
  spectroscopy of entanglement {Hamiltonians}},\ }\href
  {https://doi.org/10.1038/s41567-018-0151-7} {\bibfield  {journal} {\bibinfo
  {journal} {Nat. Phys.}\ }\textbf {\bibinfo {volume} {14}},\ \bibinfo {pages}
  {827} (\bibinfo {year} {2018})}\BibitemShut {NoStop}%
\bibitem [{\citenamefont {Giudici}\ \emph {et~al.}(2018)\citenamefont
  {Giudici}, \citenamefont {Mendes-Santos}, \citenamefont {Calabrese},\ and\
  \citenamefont {Dalmonte}}]{GMSCD18}%
  \BibitemOpen
  \bibfield  {author} {\bibinfo {author} {\bibfnamefont {G.}~\bibnamefont
  {Giudici}}, \bibinfo {author} {\bibfnamefont {T.}~\bibnamefont
  {Mendes-Santos}}, \bibinfo {author} {\bibfnamefont {P.}~\bibnamefont
  {Calabrese}},\ and\ \bibinfo {author} {\bibfnamefont {M.}~\bibnamefont
  {Dalmonte}},\ }\bibfield  {title} {\bibinfo {title} {Entanglement
  {Hamiltonians} of lattice models via the {Bisognano-Wichmann} theorem},\
  }\href {https://doi.org/10.1103/PhysRevB.98.134403} {\bibfield  {journal}
  {\bibinfo  {journal} {Phys. Rev. B}\ }\textbf {\bibinfo {volume} {98}},\
  \bibinfo {pages} {134403} (\bibinfo {year} {2018})}\BibitemShut {NoStop}%
\bibitem [{\citenamefont {Mendes-Santos}\ \emph {et~al.}(2019)\citenamefont
  {Mendes-Santos}, \citenamefont {Giudici}, \citenamefont {Dalmonte},\ and\
  \citenamefont {Rajabpour}}]{MSGDR19}%
  \BibitemOpen
  \bibfield  {author} {\bibinfo {author} {\bibfnamefont {T.}~\bibnamefont
  {Mendes-Santos}}, \bibinfo {author} {\bibfnamefont {G.}~\bibnamefont
  {Giudici}}, \bibinfo {author} {\bibfnamefont {M.}~\bibnamefont {Dalmonte}},\
  and\ \bibinfo {author} {\bibfnamefont {M.~A.}\ \bibnamefont {Rajabpour}},\
  }\bibfield  {title} {\bibinfo {title} {Entanglement {Hamiltonian} of quantum
  critical chains and conformal field theories},\ }\href
  {https://doi.org/10.1103/PhysRevB.100.155122} {\bibfield  {journal} {\bibinfo
   {journal} {Phys. Rev. B}\ }\textbf {\bibinfo {volume} {100}},\ \bibinfo
  {pages} {155122} (\bibinfo {year} {2019})}\BibitemShut {NoStop}%
\end{thebibliography}%

\end{document}